\documentclass[10pt]{revtex4}
\usepackage{amssymb,amsmath}
\usepackage{latexsym}
\usepackage{color,paralist}
\usepackage[normalem]{ulem}
\usepackage{suffix}
\usepackage{epsfig}

\DeclareMathOperator{\Tr}{Tr}
\DeclareMathOperator{\tr}{tr}

\newcommand\VTT[1]{\text{{\color[rgb]{1,0,0}~[#1]}}}
\WithSuffix\newcommand\VTT*[1]{{{\color[rgb]{1,0,0}[#1]}}}
\WithSuffix\newcommand\VTT+[1]{{{\color[rgb]{1,0,0}[#1]\\}}}
\WithSuffix\newcommand\VTT@[1]{{{\color[rgb]{1,0,0}[#1]\\\\}}}

\definecolor{mysquiggly}{rgb}{0.9,0,0}

\makeatletter
\def\squiggly{\bgroup \markoverwith{\textcolor{mysquiggly}{\lower3.5\p@\hbox{{\sixly \char58}}}}\ULon}
\makeatother

\allowdisplaybreaks

\begin{document}

\title{ MOG without anomaly }

\author{Alireza Sepehri}
\email{alireza.sepehri3@gmail.com} \affiliation{
Research Institute for Astronomy and Astrophysics of Maragha (RIAAM), P.O. Box 55134-441, Maragha, Iran}

\author{Tooraj Ghaffary }
\email{ghaffary@iaushiraz.ac.ir} \affiliation{Department of Science, Shiraz Branch, Islamic Azad University, Shiraz, Iran}

\author{Yaghoob Naimi}
\email{yaghoob.naimi@gmail.com} \affiliation{Department of Physics, Lamerd Higher Education Center, Lamerd, Iran}

\begin{abstract}
We obtain the action of Moffat's Modified Gravity (MOG), a scalar-tensor-vector theory of gravitation, by generalizing the Horava-Witten mechanism to fourteen dimensions. We show that the resulting theory is anomaly-free. We propose an extended version of MOG that includes fermionic fields.


 \end{abstract}
 \date{\today}

\maketitle
\section{Introduction}

Moffat's proposed modified gravity theory (MOG) is a scalar-tensor-vector theory of gravitation: in addition to the tensor field of General Relativity, it also contains scalar and vector fields \cite{p1}. The two scalar fields and the massive Proca vector field of this theory add several new degrees of freedom to the system, which can help us consider the evolution of universe in new ways \cite{p2}. This theory also offers a way to investigate the origin of gravitational waves, which are emitted by merging black holes and detected by the LIGO-Virgo collaboration data \cite{p3}. The question then arises: what is the origin of this theory? Also, is this gravitational model anomaly-free? We attempt to answer these questions by using the Horava-Witten mechanism.

 In 1995, Horava and Witten have shown that in eleven dimensions, all anomalies in field theory and supergravity can be canceled \cite{b1,b2}. In their model, which is known as M-theory, the 10-dimensional $E_{8}\times E_{8}$ heterotic supergravity is generalized to an 11-dimensional supergavity theory on the orbifold $R^{10}\times S^{1}/Z_{2}$ and its anomaly is canceled. In 1996, in order to solve the cosmological constant problem in four dimensions, Vafa suggested an extension of Witten's proposal to twelve dimensions. He reformulated the type IIB theory in terms of a 12-dimensional ``F-theory'' and showed that compactification of M-theory on a manifold ``$K$'', which admits an eliptic fibration, is  equivalent to compactification of F-theory on Calabi-Yau threefolds \cite{b3}. Until 2006, the algebra that could apply in M-theory and F-theory and produce the expected actions for branes was unclear. About ten years ago, Bagger and his co-authors introduced Lie three-algebra and formulated all Lagrangians in terms of it \cite{b4,b9,b10,b11}. However, we have no exact information about the structure of the world and its exact dimensionality. In fact, we cannot even limit it to eleven or twelve dimensions. For this reason, supergravity in twelve dimensions were considered and its solutions have been obtained \cite{b7,b8}.  We will generalize these mechanisms to fourteen dimensions, remove the anomalies and obtain the exact form of action for MOG.

This paper consists of two main parts. In section \ref{o1}, we show that by generalizing the Horava-Witten mechanism to fourteen dimensions, the action MOG emerges and the anomaly is removed.  In section \ref{o2}, we extend this gravity theory by including a fermionic field.

\section{MOG without anomaly in fourteen dimensions }\label{o1}

Our goal is to show that by adding a 3-dimensional manifold to 11-dimensional spacetime in the Horava-Witten mechanism, all anomalies can be removed and an action without anomaly can be produced. This action is identical to the action of the modified gravity (MOG) theory presented in \cite{p1}.

First, we introduce the Horava-Witten mechanism in  11-dimensional spacetime. In this model, the bosonic part of the action in 11-dimensional supergravity (SUGRA) is given by \cite{b1,b2}:

     \begin{eqnarray}
     S_{\rm Bosonic-SUGRA}&=& \frac{1}{\bar{\kappa}^{2}}\int d^{11}x\sqrt{g}\Big(-\frac{1}{2}R-\frac{1}{48}G_{IJKL}G^{IJKL}\Big) + S_{\rm CGG}, \nonumber\\
     S_{\rm CGG}&=&-\frac{\sqrt{2}}{3456\bar{\kappa}^{2}}\int_{M^{11}}d^{11}x \varepsilon^{I_{1}I_{2}...I_{11}}C_{I_{1}I_{2}I_{3}}G_{I_{4}...I_{7}}G_{I_{8}...I_{11}}, \label{s1}
     \end{eqnarray}
where $\varepsilon^{I_1I_2..I_k}$ is the rank-$k$ Levi-Civita pseudotensor  and CGG is used to denote the product term of the three-form field $C_{I_{1}I_{2}I_{3}}$ and four-form field $G_{IJKL}$, which are directly related to the gauge field $A^I$, field strength $F^{IJ}$ and Ricci curvature $R^{IJ}$ \cite{b2}:

     \begin{eqnarray}
      G_{IJKL}&=&-\frac{3}{\sqrt{2}}\frac{\kappa^{2}}{\lambda^{2}}\varepsilon(x^{11})(F_{[IJ}F_{KL]}-R_{[IJ}R_{KL]})+...,\nonumber\\
      \delta C_{ABC}&=&-\frac{\kappa^{2}}{6\sqrt{2}\lambda^{2}}\delta (x^{11})\tr(\epsilon_C F_{AB}-\epsilon_C R_{AB}),\nonumber\\
      G_{11ABC}&=&(\partial_{11}C_{ABC}\pm \text{23 permutations of the indices $11$ and $ABC$})+\frac{\kappa^{2}}{\sqrt{2}\lambda^{2}}\delta (x^{11})\omega_{ABC},\nonumber\\
      \delta \omega_{ABC}&=&\partial_{A}\tr(\epsilon F_{BC})+ \text{cyclic permutations of}~ABC,\nonumber\\
      F^{IJ}&=&\partial^{I}A^{J}-\partial^{J}A^{I},\nonumber\\
      R_{IJ}&=&\partial_{I}\Gamma^{B}_{JB}-\partial_{J}\Gamma^{B}_{IB} +\Gamma^{A}_{JB}\Gamma^{B}_{IA} -\Gamma^{A}_{IB}\Gamma^{B}_{JA},\nonumber\\
      \Gamma_{IJK}&=&\partial_{I}g_{JK}+\partial_{K}g_{IJ}-\partial_{J}g_{IK}, \nonumber\\
      \hat{G}_{IJ}&=&R_{IJ}-\frac{1}{2}R g_{IJ},\label{s2}
      \end{eqnarray}
where $\epsilon$ and $\epsilon_C$ characterize infinitesimal gauge transformations \cite{b2}. Here,  $\varepsilon(x^{11})$ is 1 for $x^{11}> 0$ and $-1$ for $x^{11}< 0$ and also $\delta(x^{11})=\frac{1}{2}\partial \varepsilon(x^{11})/\partial x^{11}$ is the Dirac delta function. As usual \cite{b2}, $\tr$ is 1/30th of the trace $\Tr$ in the adjoint representation for $E_8\times E_8$. The ellipsis ($...$) denotes terms that are regular near $x^{11}=0$ hence vanish there \cite{b2}.

The gauge variation of the 
CGG term in the action yields the following equation \cite{b2}:

      \begin{eqnarray}
        \delta S_{\rm CGG}|_{11}&=&-\frac{\sqrt{2}}{3456\bar{\kappa}^{2}}\int_{M^{11}}d^{11}x \varepsilon^{I_{1}I_{2}...I_{11}}\delta C_{I_{1}I_{2}I_{3}}G_{I_{4}...I_{7}}G_{I_{8}...I_{11}}\nonumber\\
        &\approx & - \frac{\bar{\kappa}^{4}}{128 \lambda^{6}}\int_{M^{10}}\Sigma_{n=1}^{5}(\tr F^{n}-\tr R^{n}+\tr(F^{n}R^{5-n})),\label{s3}
        \end{eqnarray}
%
      where $\tr X^{n}=\tr(X_{[I_{1}I_{2}}..X_{I_{2n-1}I_{2n}]})=\varepsilon^{I_{1}I_{2}..I_{2n-1}I_{2n}}X_{I_{1}I_{2}}..X_{I_{2n-1}I_{2n}}$. The above terms cancel the  anomaly of  ($S_{\rm Bosonic-SUGRA}$) in 11-dimensional manifolds \cite{b2}:

      \begin{eqnarray}
          && \delta S_{\rm CGG}|_{11}=
          -\delta S^{\rm anomaly}_{\rm Bosonic-SUGRA}.\label{ss3}
          \end{eqnarray}

       Thus, $S_{\rm CGG}$ is necessary for anomaly cancelation. Our goal now is to find a good rationale for its inclusion. We also answer  the issue of the origin of CGG terms in 11-dimensional supergravity. In fact, we propose a theory in which CGG terms appear in the supergravity action without being added by hand. To this end, we  will show that first, there are only point like manifolds with scalars which attach to them. By joining these manifolds, 1-dimensional manifolds are emerged which gauge fields and gravitons live on them. Then, these manifolds glue to each other and build higher N-dimensional manifolds with various orders of gauge fields and curvatures. Gauge fields are strings with two ends which end produces one indice and totally field strength of gauge field has two indices. Some gauge fields join to each other and form G-fields. These G-fields are constructed from linking two strings and  have four ends. Each end of string produce one indice and  thus G-fields have four indices. By breaking one N-dimensional manifold, two lower dimensional manifolds (child manifolds) are produced which are connected by an extra manifold, called Chern-Simons manifold. Some strings are strengthed between child manifolds and produce anomaly.  At this stage, one end of some G-fields is located on Chern-Simons manifold and three other ends are placed on one of child manifolds. An observer that lives on one of child manifolds sees only three ends of some of G-fields. These fields play the role of Chern-Simons fields or C-fields in supergravity. The anomaly which is produced in child manifolds can be removed by extra terms which are produced by Chern-Simons manifold and gravity can be anomaly free. In fact, if we sum over energies of child manifolds and Chern-Simons one get the energy of initial big Manifold which is anomaly free.  We discuss this subject in detail. 
       
       Before discussing the process of formation of various manifolds with different dimensions, we should obtain a relation between string fields and matters and gravity. This helps us to re-formulate field theory in terms of derivatives of strings. To do this, first, we assume that our universe is born on a D3-brane. Thus, evolution of universe can be controlled by evolution of this brane and strings which live on it. Let us to introduce the action of D3-brane which is given by \cite{D3}:
       
       \begin{eqnarray}
       S_{D3}&=& -T_{D3}\int d^{4}y \sqrt{-det(\bar{\gamma}_{ab}+2\pi l_{s}^{2}F_{ab})}, \nonumber\\
       \bar{\gamma}_{ab}&=&g_{\mu\nu}\partial_{a}X^{\mu}\partial_{b}X^{\nu}, \nonumber\\ F_{ab} &=& \partial_{a}A_{b}-\partial_{b}A_{a} \label{D1}
       \end{eqnarray}
       
       where $A_{b}$ is the gauge field, $F_{ab}$ is the field strength, $X^{\mu}$ is the string, $g_{\mu\nu}$ is the metric, $T_{D3}$ is tension and $l_{s}$ is the string length. Substituting $X^{0}=t$ and doing some mathematical calculations, the acion D3-brane in equation (\ref{D1}) is given by \cite{D3}:
       
       \begin{eqnarray}
       S_{D3}&=& -T_{D3}\int d^{4}y \sqrt{1+ g_{ij}\partial_{a}X^{i}\partial^{a}X^{j} -4\pi^{2} l_{s}^{4}F_{ab}F^{ab}},  \label{D2}
       \end{eqnarray}
       
       To construct our universe on a D3-brane, this action should be equal to the action of fields and gravity in 4-dimensional universe. The action of matter and gravity is given by:
       
       \begin{eqnarray}
       S_{Gravity-Matter}&=& \int d^{4}y \sqrt{-g}\Big(R +g_{ab}\partial^{a}\phi\partial^{b}\phi -i\bar{\psi}\gamma^{a}\partial_{a}\psi+A_{i}A^{i}+\frac{1}{2}\phi^{2}+1\Big),  \label{D3}
       \end{eqnarray}
       
       where $\phi$ is the scalar field and $\psi$ is the fermionic field. Puting equation of (\ref{D2}) equal to the equation (\ref{D3}) and doing some mathematical calculations, we obtain the relation between strings and matter fields:

       \begin{eqnarray}
       S_{Gravity-Matter}&=& S_{D3}  \nonumber\\ \Longrightarrow X^{i} &=& \int dy^{i} \sqrt{-1 + 4\pi^{2} l_{s}^{4}F_{ij}F^{ij} + \Big(\sqrt{-g}(1+g^{ij}R_{ij}+A_{i}A^{i}+\frac{1}{2}\phi^{2} +g_{ij}\partial^{i}\phi\partial^{j}\phi -i\bar{\psi}\gamma^{i}\partial_{i}\psi)\Big)^{2}},  \label{D4}
       \end{eqnarray}
       
       where we have assumed $T_{D3}\simeq 1$ and $4\pi^{2} l_{s}^{4}\simeq 1$. This equation shows that there is direct relation between strings and fields in 4-dimensional field theory. Using this relation, we can consider the process of formation of manifolds and fields which live on them. 
       
  At this stage, we will show that at the beginning, there are point like manifolds in space (See Figure 1) which strings are attached to them. These manifolds have only one dimension in direction of time. All interactions between  strings  on one manifold are the same and are concentrated on one point which manifold is located on it. The potential of these interactions can be shown by a delta function and thus, the energy of manifold tends to one.

   \begin{eqnarray}
&& V(\tilde{X}^{I})= \delta(\tilde{X}^{I}) \nonumber\\&&  E_{M^{0}}=1=\int_{M^{0}} d\tilde{X}^{I} V(\tilde{X}^{I})=\int_{M^{0}} d\tilde{X}^{I} \delta(\tilde{X}^{I})= \nonumber\\&&  \int_{M^{0}} d\tilde{X}^{I} (\frac{1}{\sqrt{2\pi y}}e^{-\frac{\tilde{X}^{I}\tilde{X}_{I}}{2y}})	\label{spm1}
  \end{eqnarray} 
  
  where $M^{0}$  denotes the point like manifold, $\tilde{X}^{I}$'s are strings which attached to them and $y$ is the length of point which shrinks to zero. With new redifinition of string fields  $\tilde{X}^{I}\longrightarrow \sqrt{2\pi y} X^{I}$, we get:

  \begin{eqnarray}
  && E_{M^{0}}=1= \int_{M^{0}} dX^{I} e^{-\pi X^{I}X_{I}}	\label{spm2}
  \end{eqnarray} 
  
   We can calculate the integral and obtain a solution for strings ($X^{I}$):

   \begin{eqnarray}
   && \int_{M^{0}} dX^{I} e^{-\pi X^{I}X_{I}}=1\longrightarrow \nonumber\\&&  \frac{1}{2}erf(\sqrt{ X^{I}X_{I}\pi})=1 \longrightarrow \nonumber\\&& X^{I}\approx \epsilon^{J}I\label{spm3}
   \end{eqnarray} 
   
     where I is the unitary matrix. This equation shows that at the beginning, there is no interaction between strings and they are the same. In fact, there is a very high symmetry for the early stages of world and all matters have the same origin.

     \begin{figure*}[thbp]
     	\begin{center}
     		\begin{tabular}{rl}
     			\includegraphics[width=8cm]{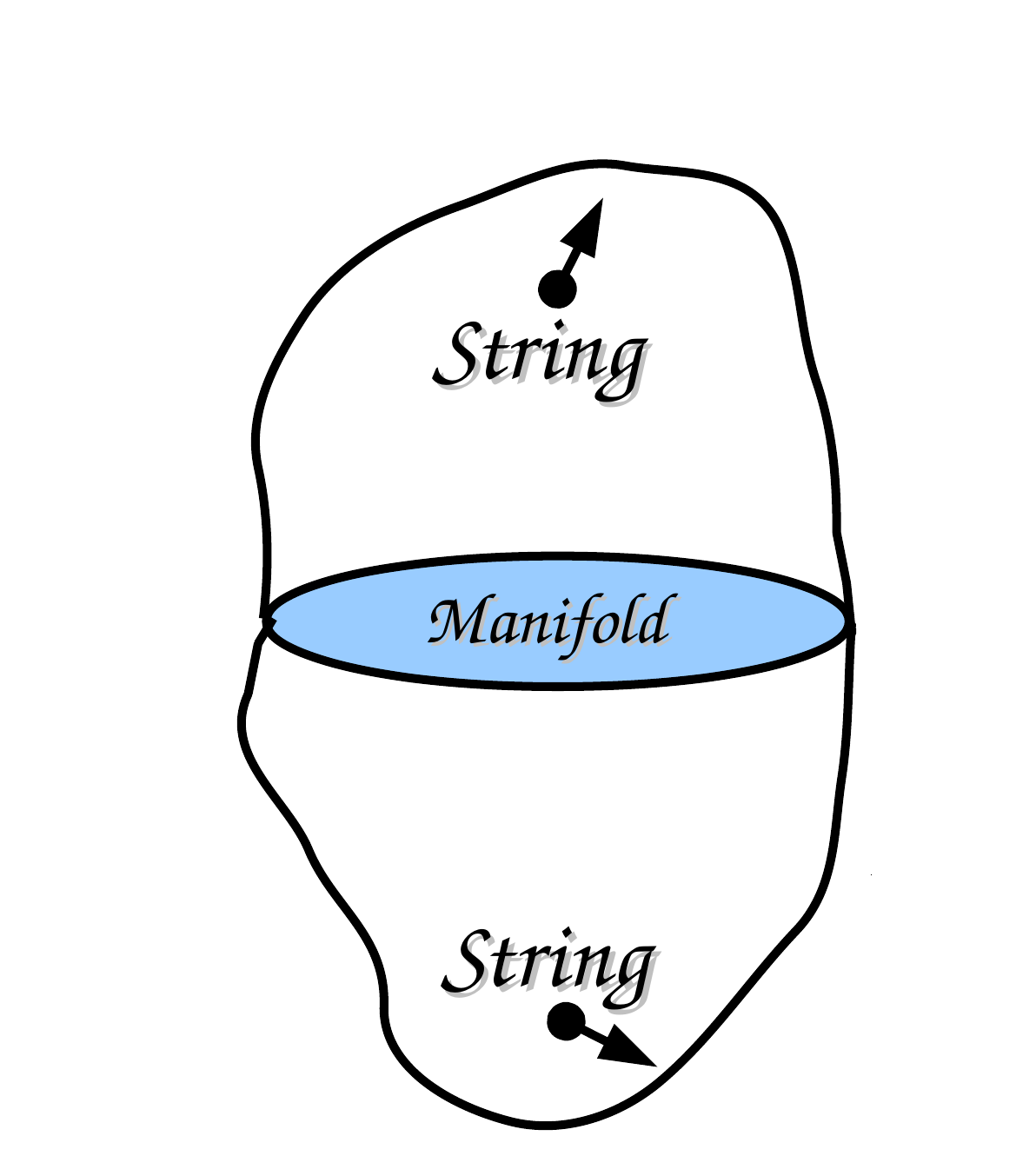}
     		\end{tabular}
     	\end{center}
     	\caption{  Point-like manifolds and attached strings. }
     \end{figure*}
     
     At second stage, two point like manifolds join to each other and form a one dimensional manifold (See Figure 2.). The interactions between strings on each point like manifold can be explained by a delta function and total interaction of strings can be described by a big delta function.  Because, all strings which are attached on  one point like manifold, interact with each other on  one point and thus total potential is  zero in other points and infinite  in this special point. Integrating over all these potentals yields:

     \begin{eqnarray}
     && E_{M_{1}^{0}+M_{2}^{0}}=1=\int_{M_{1}^{0}+M_{2}^{0}} d\tilde{X_{1}}^{I}d\tilde{X_{2}}_{I} \delta(\tilde{X_{1}}^{I}\tilde{X_{2}}^{I})= \nonumber\\&&\int_{M_{1}^{0}+M_{2}^{0}} d\tilde{X_{1}}^{I}d\tilde{X_{2}}_{I} \delta(\tilde{X_{1}}^{I})\delta(\tilde{X_{2}}^{I})= \nonumber\\&&  \int_{M_{1}^{0}+M_{2}^{0}} d\tilde{X_{1}}^{I}d\tilde{X_{2}}_{I} (\frac{1}{\sqrt{2\pi y_{1}}}e^{-\frac{\tilde{X_{1}}^{I}\tilde{X_{1}}_{I}}{2y_{1}}})(\frac{1}{\sqrt{2\pi y_{2}}}e^{-\frac{\tilde{X_{2}}^{I}\tilde{X_{2}}_{I}}{2y_{2}}})	\label{spm4}
     \end{eqnarray}

     where $M_{1/2}^{0}$  denote the first or second point like manifolds, $\tilde{X_{1/2}}^{I}$'s are strings which attached to them and $y_{1/2}$ are the length of points which shrinks to zero. In above equation,  there are two integrations which one is related to time and other is corresponded to space. For this reason, we have 1-dimensional manifold in space and 2-dimensional manifold in space-time. Because of special properties of time, we only regard dimension of manifold in space.  Again, we  redifine  string fields  $\tilde{X}^{I}\longrightarrow \sqrt{2\pi y} X^{I}$ and obtain:

     \begin{eqnarray}
     && E_{M^{1}}=1= \int_{M_{1}^{0}+M_{2}^{0}} dX_{1}^{I}dX_{2 I} e^{-\pi X_{1}^{I}X_{1 I}}e^{-\pi X_{2}^{I}X_{2 I}}	\label{spm5}
     \end{eqnarray}

      When, point like manifolds join to each other and construct a one dimensional manifold,  strings will be  functions of coordinates of both manifolds ($X^{I}(y_{1},y_{2})$). We can rewrite equation (\ref{spm5}) as:

      \begin{eqnarray}
      && E_{M^{1}}=1= \nonumber\\&& \int_{M_{1}^{0}+M_{2}^{0}}dy_{1}^{I}dy_{2 I} (\frac{dX_{1}^{I}}{dy_{1}^{I}}\frac{dX_{2 I}}{dy_{2 I}}+\frac{dX_{1}^{I}}{dy_{2 I}}\frac{dX_{2 I}}{dy_{1}^{I}}) e^{-\pi X_{1}^{I}X_{1 I}}e^{-\pi X_{2}^{I}X_{2 I}} +	\nonumber\\&& \int_{M_{1}^{0}+M_{2}^{0}}dy_{1}^{I}dy_{1 I} (\frac{dX_{1}^{I}}{dy_{1}^{I}}\frac{dX_{2 I}}{dy_{1 I}}) e^{-\pi X_{1}^{I}X_{1 I}}e^{-\pi X_{2}^{I}X_{2 I}}+\nonumber\\&& \int_{M_{1}^{0}+M_{2}^{0}}dy_{2}^{I}dy_{2 I} (\frac{dX_{1}^{I}}{dy_{2}^{I}}\frac{dX_{2 I}}{dy_{2 I}}) e^{-\pi X_{1}^{I}X_{1 I}}e^{-\pi X_{2}^{I}X_{2 I}}\label{spm6}
      \end{eqnarray}

   In above equation, one of coordinates can be known as time coordinate and other can be known as the space-coordinate. Thus, ignoring time direction, above equation explains the energy along one dimensional manifold in space. By using taylor method, we expand exponential functions over the crossed points. We obtain:

  \begin{eqnarray}
  && E_{M^{1}}=1= \nonumber\\&& \int_{M_{1}^{0}+M_{2}^{0}}dy_{1}^{I}dy_{2 I} (\frac{dX_{1}^{I}}{dy_{1}^{I}}\frac{dX_{2 I}}{dy_{2 I}}+\frac{dX_{1}^{I}}{dy_{2 I}}\frac{dX_{2 I}}{dy_{1}^{I}}) +\nonumber\\&& \int_{M_{1}^{0}+M_{2}^{0}}dy_{1}^{I}dy_{1 I} (\frac{dX_{1}^{I}}{dy_{1}^{I}}\frac{dX_{2 I}}{dy_{1 I}}) +\int_{M_{1}^{0}+M_{2}^{0}}dy_{2}^{I}dy_{2 I} (\frac{dX_{1}^{I}}{dy_{2}^{I}}\frac{dX_{2 I}}{dy_{2 I}})+\nonumber\\&& \int_{M_{1}^{0}+M_{2}^{0}}dy_{1}^{I}dy_{2 I} (\frac{dX_{1}^{I}}{dy_{1}^{I}}\frac{dX_{2 I}}{dy_{2 I}}+\frac{dX_{1}^{I}}{dy_{2 I}}\frac{dX_{2 I}}{dy_{1}^{I}}) \pi^{2} \frac{\partial}{\partial y_{2 I}}\frac{\partial}{\partial y_{1}^{I}}(X_{1}^{I}X_{1 I}) \frac{\partial}{\partial y_{1 I}}\frac{\partial}{\partial y_{2}^{I}}( X_{2}^{I}X_{2 I})(y_{2}^{I}-y_{0}^{I})(y_{1 I}-y_{0 I}) +	\nonumber\\&& \int_{M_{1}^{0}+M_{2}^{0}}dy_{1}^{I}dy_{1 I} (\frac{dX_{1}^{I}}{dy_{1}^{I}}\frac{dX_{2 I}}{dy_{1 I}}) \pi^{2} \frac{\partial}{\partial y_{1 I}}\frac{\partial}{\partial y_{1}^{I}}(X_{1}^{I}X_{1 I}) \frac{\partial}{\partial y_{1 I}}\frac{\partial}{\partial y_{1}^{I}}( X_{2}^{I}X_{2 I})(y_{1}^{I}-y_{0}^{I})(y_{1 I}-y_{0 I})+\nonumber\\&& \int_{M_{1}^{0}+M_{2}^{0}}dy_{2}^{I}dy_{2 I} (\frac{dX_{1}^{I}}{dy_{2}^{I}}\frac{dX_{2 I}}{dy_{2 I}}) \pi^{2} \frac{\partial}{\partial y_{2 I}}\frac{\partial}{\partial y_{2}^{I}}(X_{1}^{I}X_{1 I}) \frac{\partial}{\partial y_{2 I}}\frac{\partial}{\partial y_{2}^{I}}( X_{2}^{I}X_{2 I})(y_{2}^{I}-y_{0}^{I})(y_{2 I}-y_{0 I})-	\nonumber\\&& \int_{M_{1}^{0}+M_{2}^{0}}dy_{1}^{I}dy_{1 I} (\frac{dX_{1}^{I}}{dy_{1}^{I}}\frac{dX_{2 I}}{dy_{1 I}}) \pi^{2} \frac{\partial}{\partial y_{1}^{I}}\frac{\partial}{\partial y_{1 I}}\frac{\partial}{\partial y_{1}^{I}}(X_{1}^{I}X_{1 I}) \frac{\partial}{\partial y_{1 I}}\frac{\partial}{\partial y_{1 I}}\frac{\partial}{\partial y_{1}^{I}}( X_{2}^{I}X_{2 I})(y_{1}^{I}-y_{0}^{I})(y_{1 I}-y_{0 I})-\nonumber\\&& \int_{M_{1}^{0}+M_{2}^{0}}dy_{2}^{I}dy_{2 I} (\frac{dX_{1}^{I}}{dy_{2}^{I}}\frac{dX_{2 I}}{dy_{2 I}}) \pi^{2} \frac{\partial}{\partial y_{2 I}}\frac{\partial}{\partial y_{2 I}}\frac{\partial}{\partial y_{2}^{I}}(X_{1}^{I}X_{1 I}) \frac{\partial}{\partial y_{2}^{I}}\frac{\partial}{\partial y_{2 I}}\frac{\partial}{\partial y_{2}^{I}}( X_{2}^{I}X_{2 I})(y_{2}^{I}-y_{0}^{I})(y_{2 I}-y_{0 I})\label{spm7}
  \end{eqnarray} 
  
  In above equation, any change in the shape of strings is shown by an special derivative respect to coordinates. These different shapes produce different types of matters like scalars, spinors and gauge fields. For example, symmetric derivatives respect to coordinates, produce gravitons and anti-symmetric derivatives produce gauge fields (See figures 3, 4 and 5). Using the result in equation (\ref{D4}), we obtain:

  \begin{eqnarray}
  && X_{1}^{I}X_{1 I}=X_{2}^{I}X_{2 I} \nonumber\\&& \nonumber\\&& \nonumber\\&&  \frac{dX_{i}^{I}}{dy_{j}^{I}}=e^{i}_{j} \quad i,j=1,2\nonumber\\&& A^{1}_{I}\rightarrow \frac{\partial}{\partial y_{1}^{I}}(X_{1}^{I}X_{1 I}) \nonumber\\&& F_{12}\rightarrow  \Big( \frac{\partial}{\partial y_{2 I}}\frac{\partial}{\partial y_{1}^{I}}(X_{1}^{I}X_{1 I})- \frac{\partial}{\partial y_{1 I}}\frac{\partial}{\partial y_{2}^{I}}(X_{1}^{I}X_{1 I})\Big) \nonumber\\&& g_{12}\rightarrow  \Big( \frac{\partial}{\partial y_{2 I}}\frac{\partial}{\partial y_{1}^{I}}(X_{1}^{I}X_{1 I})+ \frac{\partial}{\partial y_{1 I}}\frac{\partial}{\partial y_{2}^{I}}(X_{1}^{I}X_{1 I})\Big)=\nonumber\\&& \Big( \frac{\partial X_{1}^{I}}{\partial y_{2 I}}\frac{\partial X_{1 I}}{\partial y_{1}^{I}}+ \frac{\partial X_{1}^{I}}{\partial y_{1 I}}\frac{\partial X_{1 I}}{\partial y_{2}^{I}}+...\Big)= \nonumber\\&& \Big( e^{1}_{2}e_{1 1}+e_{1 1}e^{1}_{2}\Big)+... \nonumber\\&& \nonumber\\&& \nonumber\\&& \frac{\partial}{\partial y_{2 I}}\frac{\partial}{\partial y_{1}^{I}}(X_{1}^{I}X_{1 I})= \nonumber\\&& \frac{1}{2}\Big( \frac{\partial}{\partial y_{2 I}}\frac{\partial}{\partial y_{1}^{I}}(X_{1}^{I}X_{1 I})+ \frac{\partial}{\partial y_{1 I}}\frac{\partial}{\partial y_{2}^{I}}(X_{1}^{I}X_{1 I})\Big)+\nonumber\\&& \frac{1}{2}\Big( \frac{\partial}{\partial y_{2 I}}\frac{\partial}{\partial y_{1}^{I}}(X_{1}^{I}X_{1 I})- \frac{\partial}{\partial y_{1 I}}\frac{\partial}{\partial y_{2}^{I}}(X_{1}^{I}X_{1 I})\Big)=\nonumber\\&& \frac{1}{2} g_{12} + \frac{1}{2} F_{12} \nonumber\\&&\nonumber\\&&\nonumber\\&& \phi \rightarrow \frac{\partial}{\partial y_{1 I}}\frac{\partial}{\partial y_{1}^{I}}(X_{1}^{I}X_{1 I})\nonumber\\&&\nonumber\\&&\nonumber\\&& \sqrt{-g}g^{12} \rightarrow (\frac{dX_{1}^{I}}{dy_{2}^{I}}\frac{dX_{2 I}}{dy_{2 I}}) \label{spm8}
  \end{eqnarray} 
  
  where $g$ is the metric, $F$ is the field strenth, $A$ is the gauge field and $\phi$ is the scalar field (See figures 3, 4 and 5). It is clear from figure 1 that scalars ($\phi$) are produced by joining two ends of strings on point like manifolds. Figure 2, shows that if one end of string is located on one point like manifold and another point is placed on another point manifold, one gauge field ($F_{ij}$) is emerged. By changing the place of i and j, direction of fields is reversed. Figure 5 indicates that near the linked points of point-like manifolds, strings form symmetric shapes which can be known as a graviton.

     We also assume that point like manifolds are very closed to each other and thus, we have:

       \begin{eqnarray}
       && y_{2}^{I}-y_{0}^{I}=y_{1}^{I}-y_{0}^{I}=\sigma\longrightarrow \frac{1}{\pi}  \label{spm9}
       \end{eqnarray}

       Using equations (\ref{spm9} and \ref{spm8}) in equation (\ref{spm7}), we obtain:

      \begin{eqnarray}
      && E_{M^{1}}=1= (X^{I}X_{I})- \int_{M_{1}^{0}+M_{2}^{0}}d^{2}y\sqrt{-g}\Big(\frac{1}{4}\partial_{i}\phi\partial^{i}\phi + \frac{1}{4}\epsilon^{ijkf} F_{ij}F_{kf} - \frac{1}{2}g^{ij}g_{ij} + V(\phi)\Big ) \quad i,j=1,2\nonumber\\&& V(\phi)=1+ \frac{\phi^{2}}{2}  \label{spm10}
      \end{eqnarray}  
       
  This equation shows that by joining point like manifolds and formation of 1-dimensional manifolds, gauge fields and scalars are appeared. Also, the shape of energy which is produced during this process, is very the same of energy of matters in usual quantum field theory.  To consider the process of creation of gravity, we go towards next steps.

       \begin{figure*}[thbp]
       	\begin{center}
       		\begin{tabular}{rl}
       			\includegraphics[width=8cm]{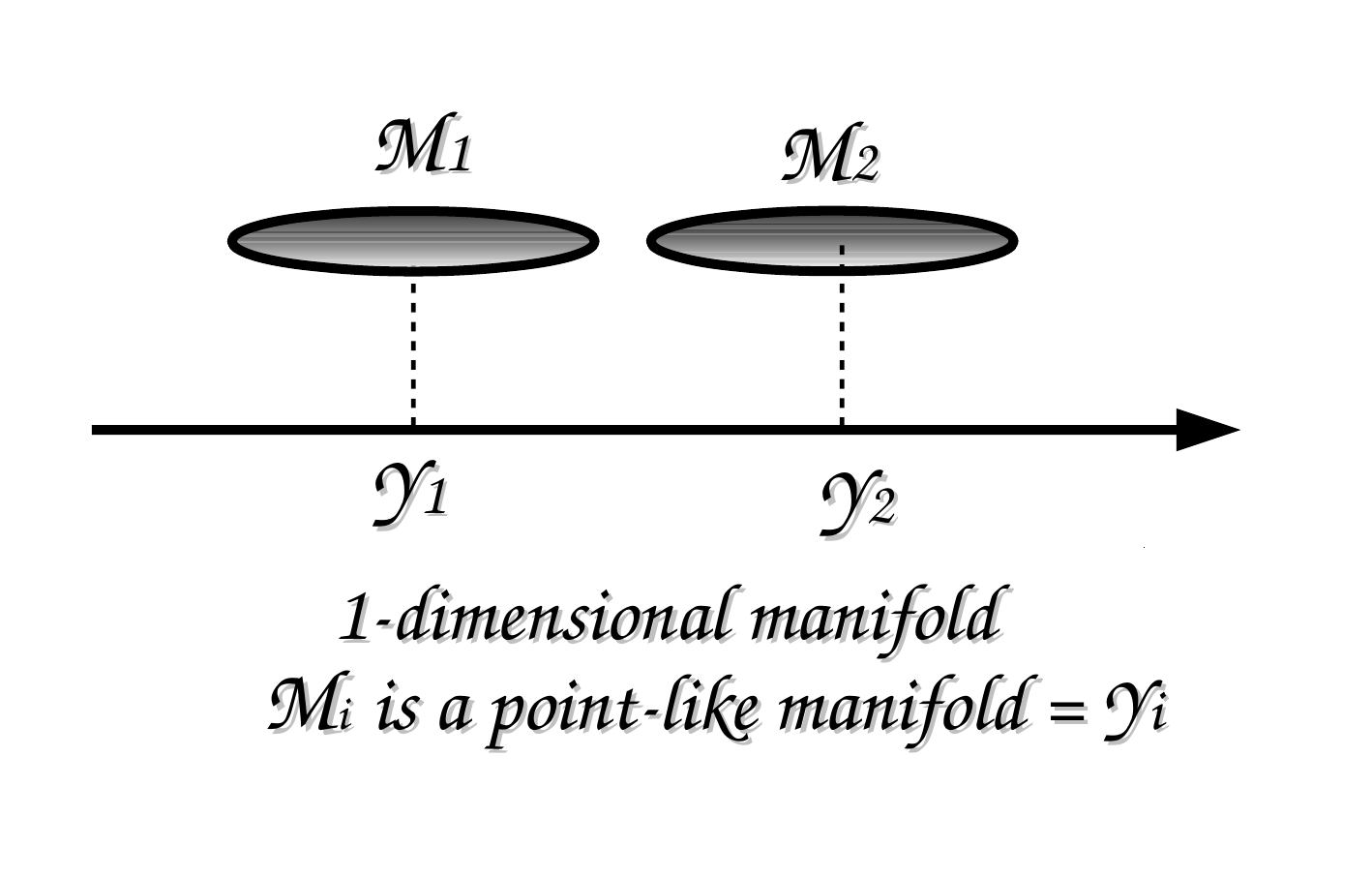}
       		\end{tabular}
       	\end{center}
       	\caption{  Constructing of one dimensional manifold by joining point like manifolds. }
       \end{figure*}

   \begin{figure*}[thbp]
   	\begin{center}
   		\begin{tabular}{rl}
   			\includegraphics[width=8cm]{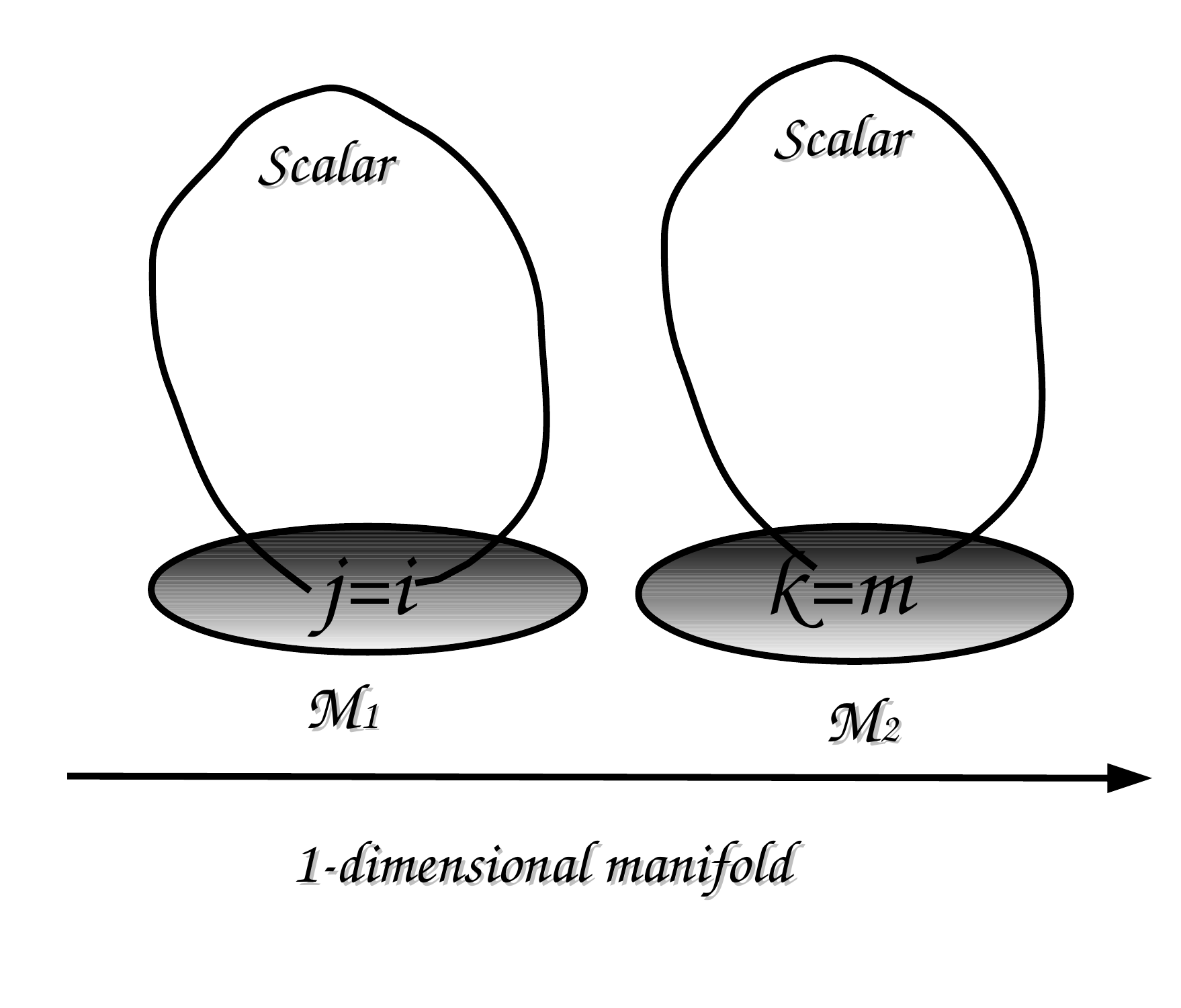}
   		\end{tabular}
   	\end{center}
   	\caption{  Strings which both ends of them is located on one point like manifold form scalar fields ($\phi$). }
   \end{figure*}
   
    \begin{figure*}[thbp]
   	\begin{center}
   		\begin{tabular}{rl}
   			\includegraphics[width=8cm]{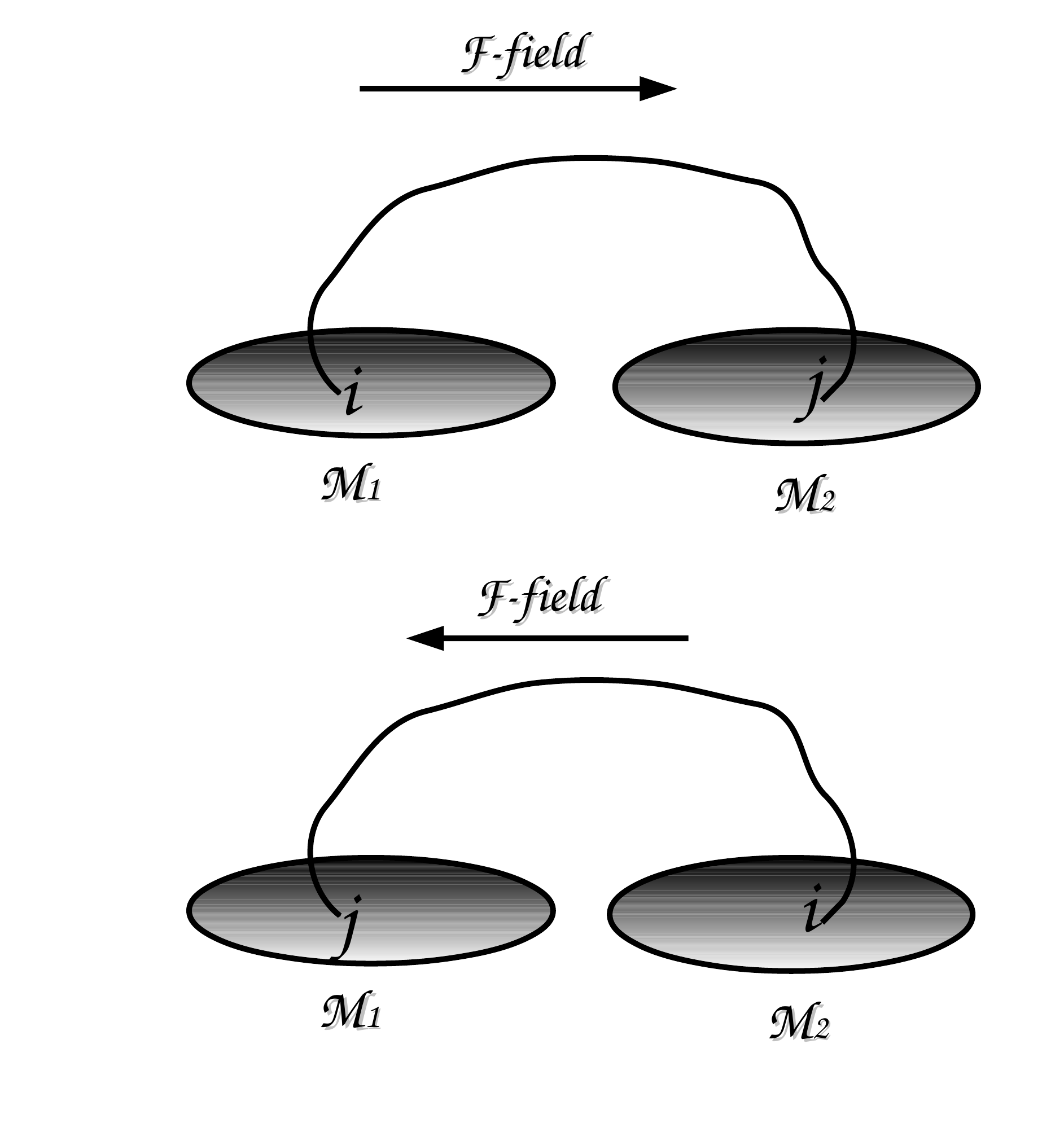}
   		\end{tabular}
   	\end{center}
   	\caption{  Strings which one end is place on one point like manifold and other is located on another point like manifold form Gauge fields (F-fields ($F_{ij}$)). }
   \end{figure*}
   
       \begin{figure*}[thbp]
       	\begin{center}
       		\begin{tabular}{rl}
       			\includegraphics[width=8cm]{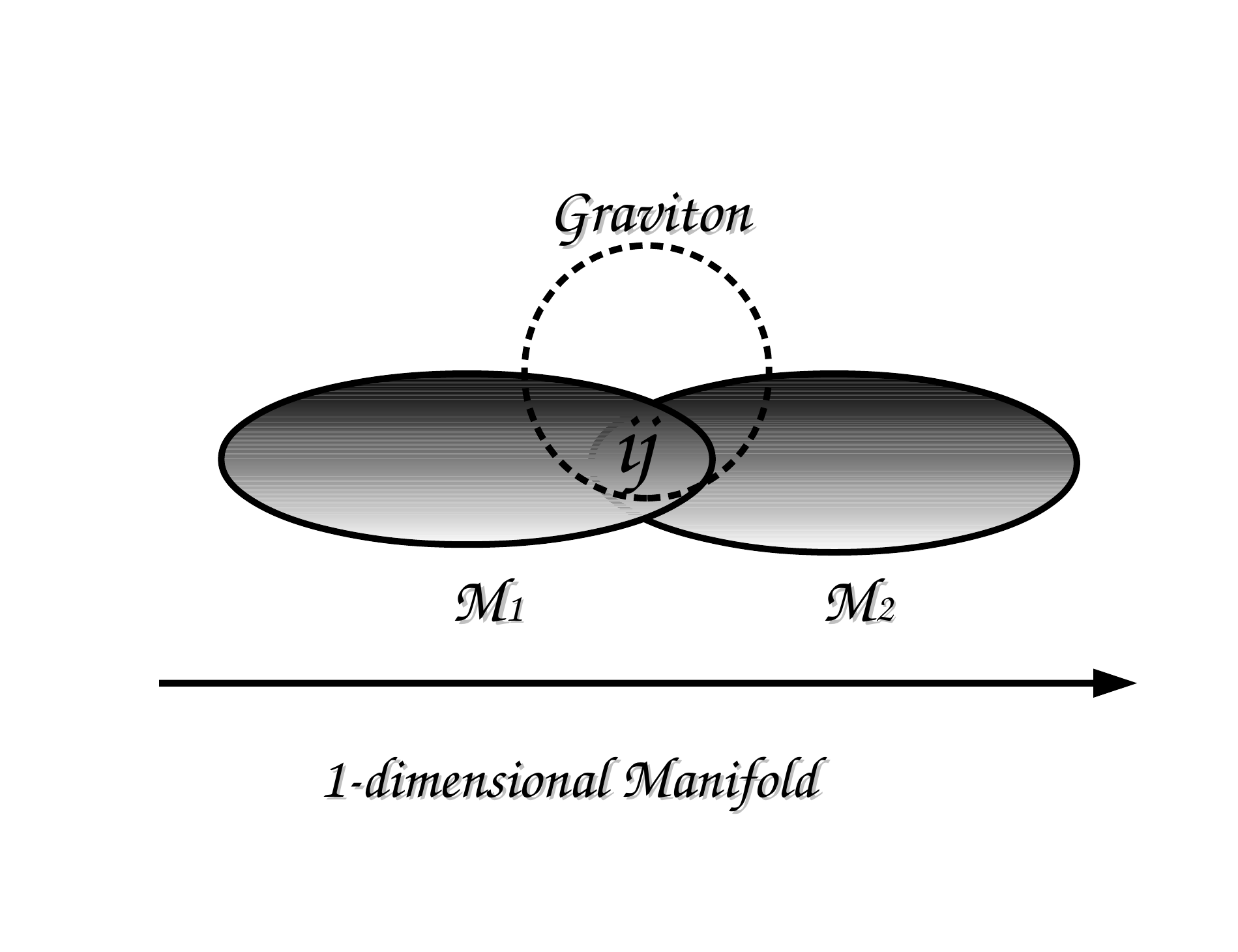}
       		\end{tabular}
       	\end{center}
       	\caption{  Strings which two ends are placed near linked point and construct a symmetric shape, form graviton fields ($g_{ij}$). }
       \end{figure*}

      At third stage, two one dimensional manifold or three point like manifolds join to each other and construct a 2-dimensional manifold (See Figure 6.). Similar to previous stages, the potential of interactions between strings on joined point like manifolds should be explained by  delta functions. We can write:

      \begin{eqnarray}
      	&& E_{M_{1}^{1}+M_{2}^{1}}=E_{M_{1}^{0}+M_{2}^{0}+M_{3}^{0}}=1=\nonumber\\&& \int_{M_{1}^{0}+M_{2}^{0}+M_{3}^{0}} d\tilde{X_{1}}^{I}d\tilde{X_{2}}_{I}d\tilde{X_{3}}_{I} \delta(\tilde{X_{1}}^{I})\delta(\tilde{X_{2}}^{I})\delta(\tilde{X_{3}}^{I})=\nonumber\\&&  \int_{M_{1}^{0}+M_{2}^{0}+M_{3}^{0}} d\tilde{X_{1}}^{I}d\tilde{X_{2}}_{I} d\tilde{X_{3}}_{I}(\frac{1}{\sqrt{2\pi y_{1}}}e^{-\frac{\tilde{X_{1}}^{I}\tilde{X_{1}}_{I}}{2y_{1}}})(\frac{1}{\sqrt{2\pi y_{2}}}e^{-\frac{\tilde{X_{2}}^{I}\tilde{X_{2}}_{I}}{2y_{2}}})(\frac{1}{\sqrt{2\pi y_{3}}}e^{-\frac{\tilde{X_{3}}^{I}\tilde{X_{3}}_{I}}{2y_{3}}})	\label{spm11}
      \end{eqnarray}
       
    Redifining  string fields  $\tilde{X}^{I}\longrightarrow \sqrt{2\pi y} X^{I}$ yields:

     \begin{eqnarray}
     	&& E_{M^{2}}=1= \int_{M_{1}^{0}+M_{2}^{0}+M_{3}^{0}} dX_{1}^{I}dX_{2 I}dX_{3 I} e^{-\pi X_{1}^{I}X_{1 I}}e^{-\pi X_{2}^{I}X_{2 I}}e^{-\pi X_{3}^{I}X_{3 I}}	\label{spm12}
     \end{eqnarray}

   By joining point like manifolds  to each other and building a 2-dimensional manifold,  strings will be  functions of coordinates of three manifolds ($X^{I}(y_{1},y_{2},y_{3})$). Thus, equation (\ref{spm12}) can be rewritten as:

   \begin{eqnarray}
   	&& E_{M^{2}}=1= \nonumber\\&& \int_{M_{1}^{0}+M_{2}^{0}+M_{3}^{0}}dy_{1}^{I}dy_{2 I}dy_{3 I} (\Sigma_{i,j,k=1,2,3}\frac{dX_{i}^{I}}{dy_{j}^{I}}\frac{dX_{j}^{I}}{dy_{k}^{I}}\frac{dX_{k}^{I}}{dy_{i}^{I}})\times \nonumber\\&& e^{-\pi X_{1}^{I}X_{1 I}}e^{-\pi X_{2}^{I}X_{2 I}}e^{-\pi X_{3}^{I}X_{3 I}} +	\nonumber\\&& \int_{M_{1}^{0}+M_{2}^{0}+M_{3}^{0}}dy_{1}^{I}dy_{1 I}dy_{1 I} (\frac{dX_{1}^{I}}{dy_{1}^{I}}\frac{dX_{2 I}}{dy_{1 I}}\frac{dX_{3 I}}{dy_{1 I}}) e^{-\pi X_{1}^{I}X_{1 I}}e^{-\pi X_{2}^{I}X_{2 I}}e^{-\pi X_{3}^{I}X_{3 I}}+\nonumber\\&&  \int_{M_{1}^{0}+M_{2}^{0}+M_{3}^{0}}dy_{2}^{I}dy_{2 I}dy_{2 I} (\frac{dX_{1}^{I}}{dy_{2}^{I}}\frac{dX_{2 I}}{dy_{2 I}}\frac{dX_{3 I}}{dy_{2 I}}) e^{-\pi X_{1}^{I}X_{1 I}}e^{-\pi X_{2}^{I}X_{2 I}}e^{-\pi X_{3}^{I}X_{3 I}}+\nonumber\\&& \int_{M_{1}^{0}+M_{2}^{0}+M_{3}^{0}}dy_{3}^{I}dy_{3 I}dy_{3 I} (\frac{dX_{1}^{I}}{dy_{3}^{I}}\frac{dX_{2 I}}{dy_{3 I}}\frac{dX_{3 I}}{dy_{3 I}}) e^{-\pi X_{1}^{I}X_{1 I}}e^{-\pi X_{2}^{I}X_{2 I}}e^{-\pi X_{3}^{I}X_{3 I}}\label{spm13}
   \end{eqnarray}

  One of these coordinates is known as time coordinate and two others are known as spacial coordinates. Ignoring time dimension, we have 2-dimensional manifold in space.  By applying taylor mechanism and expand exponential functions over the crossed points, we re-obtain all terms of equation (\ref{spm7}) in additional to following integration:

  \begin{eqnarray}
  	&& E_{M^{2}}=1= 2E_{M^{1}}-\nonumber\\&&  \int_{M_{1}^{0}+M_{2}^{0}+M_{3}^{0}}dy_{1}^{I}dy_{2 I}dy_{3 I} (\Sigma_{i,j,k=1,2,3}\frac{dX_{i}^{I}}{dy_{j}^{I}}\frac{dX_{j}^{I}}{dy_{k}^{I}}\frac{dX_{k}^{I}}{dy_{i}^{I}})\times \nonumber\\&& \pi^{3} (\Sigma_{i,j,k=1,2,3}(\frac{\partial}{\partial y_{i I}}(X_{i}^{I}X_{i I})+\frac{\partial}{\partial y_{i I}}\frac{\partial}{\partial y_{j}^{I}}(X_{i}^{I}X_{i I})+\frac{\partial}{\partial y_{i I}}\frac{\partial}{\partial y_{j}^{I}}\frac{\partial}{\partial y_{k I}}(X_{i}^{I}X_{i I})+1)\times \nonumber\\&& (1+\frac{\partial}{\partial y_{i I}}(X_{j}^{I}X_{j I})+\frac{\partial}{\partial y_{i I}}\frac{\partial}{\partial y_{j}^{I}}(X_{j}^{I}X_{j I})+ \frac{\partial}{\partial y_{j I}}\frac{\partial}{\partial y_{i}^{I}}\frac{\partial}{\partial y_{k I}}(X_{j}^{I}X_{j I}))\times \nonumber\\&& (\frac{\partial}{\partial y_{i I}}(X_{k}^{I}X_{k I})+\frac{\partial}{\partial y_{i I}}\frac{\partial}{\partial y_{j}^{I}}(X_{k}^{I}X_{k I})+\frac{\partial}{\partial y_{k I}}\frac{\partial}{\partial y_{i}^{I}}\frac{\partial}{\partial y_{j I}}(X_{k}^{I}X_{k I})+1)\times \nonumber\\&&\Big(\Sigma_{ijk=0,1,2,3}(y_{3}^{I}-y_{0}^{I})^{i}(y_{2}^{I}-y_{0}^{I})^{j}(y_{1 I}-y_{0 I})^{k}\Big) +	...\label{spm14}
  \end{eqnarray} 
 
Thus, energy of 2-dimensional manifold can be obtained by summing over energies of 1-dimensional manifolds plus extra energy which is needed for linking manifolds. Using results in equation (\ref{D4} and \ref{spm8}), we can calculate terms in above integration:

  \begin{eqnarray}
  && X_{1}^{I}X_{1 I}=X_{2}^{I}X_{2 I}=X_{3}^{I}X_{3 I} \nonumber\\&& \nonumber\\&& \nonumber\\&&    \frac{\partial}{\partial y_{3 I}}\frac{\partial}{\partial y_{2 I}}\frac{\partial}{\partial y_{1}^{I}}(X_{1}^{I}X_{1 I})= \nonumber\\&& \frac{1}{2} \frac{\partial}{\partial y_{3 I}}(g_{12}) + \frac{1}{2} \epsilon^{123}\frac{\partial}{\partial y_{3 I}}(F_{12})+...\Rightarrow \nonumber\\&&\nonumber\\&&\nonumber\\&& \Sigma_{ijk}\frac{\partial}{\partial y_{k I}}\frac{\partial}{\partial y_{j I}}\frac{\partial}{\partial y_{i}^{I}}(X_{i}^{I}X_{i I})= \nonumber\\&& \frac{1}{2}\Sigma_{ijk} (\frac{\partial}{\partial y_{k I}}(g_{ij}) +  \epsilon^{ijk}\frac{\partial}{\partial y_{k I}}(F_{ij}))+...=\nonumber\\&&\frac{1}{2}\Sigma_{ijk} (\frac{\partial}{\partial y_{k I}}(g_{ij})+\frac{\partial}{\partial y_{i I}}(g_{kj})-\frac{\partial}{\partial y_{i I}}(g_{kj}) +  \epsilon^{ijk}\frac{\partial}{\partial y_{k I}}(F_{ij}))+....=\nonumber\\&& \frac{1}{2}\Sigma_{ijk} (\epsilon_{I}^{ijk}\Gamma_{ijk} +  \epsilon^{ijk}\frac{\partial}{\partial y_{k I}}(F_{ij}))+....  \label{spm15}
  \end{eqnarray} 
  
    Similar to previous stages, we assume that point like manifolds are very closed to each other and thus, we can write:

   \begin{eqnarray}
   &&  y_{3}^{I}-y_{0}^{I}= y_{2}^{I}-y_{0}^{I}=y_{1}^{I}-y_{0}^{I}=\sigma\longrightarrow \frac{1}{\pi}  \label{spm16}
   \end{eqnarray}

   Using equations ( \ref{spm7},\ref{spm8}, \ref{spm9}, \ref{spm10}, \ref{spm15}, \ref{spm16}) in equation (\ref{spm14}), we obtain:

   \begin{eqnarray}
   && E_{M^{2}}=1= \epsilon^{I}(X^{I}X_{I}X_{I})-\nonumber\\&&  \int_{M^{2}}d^{3}y\sqrt{-g}\Big(\frac{1}{4}\partial_{i}\phi\partial^{i}\phi + \frac{1}{4}\epsilon^{ijkf} F_{ij}F_{kf} + \frac{1}{6} \epsilon^{ijkflm}F_{ij}F_{kf}F_{lm}+ \nonumber\\&&  + \frac{1}{6} \epsilon^{ijkflm}\partial_{l}F_{ij}\partial_{m}F_{kf}-\frac{1}{2}g^{ij}g_{ij} + \epsilon^{ijk}\Gamma_{ijk}+\Gamma^{ijk}\Gamma_{ijk}+ V(\phi)\Big ) \quad i,j=1,2,3\nonumber\\&& V(\phi)=1+ \frac{\phi^{2}}{2}-\frac{\phi^{3}}{6}  \label{spm17}
   \end{eqnarray}  
   
   This equation indicates that by joining 1-dimensional manifolds and constructing 2-dimensional manifolds, the first signature of gravity is appeared. In fact, energy of system not only includes energy of matters but also contains some $\Gamma$ terms which are the first steps towards gravity.
     
   \begin{figure*}[thbp]
   	\begin{center}
   		\begin{tabular}{rl}
   			\includegraphics[width=8cm]{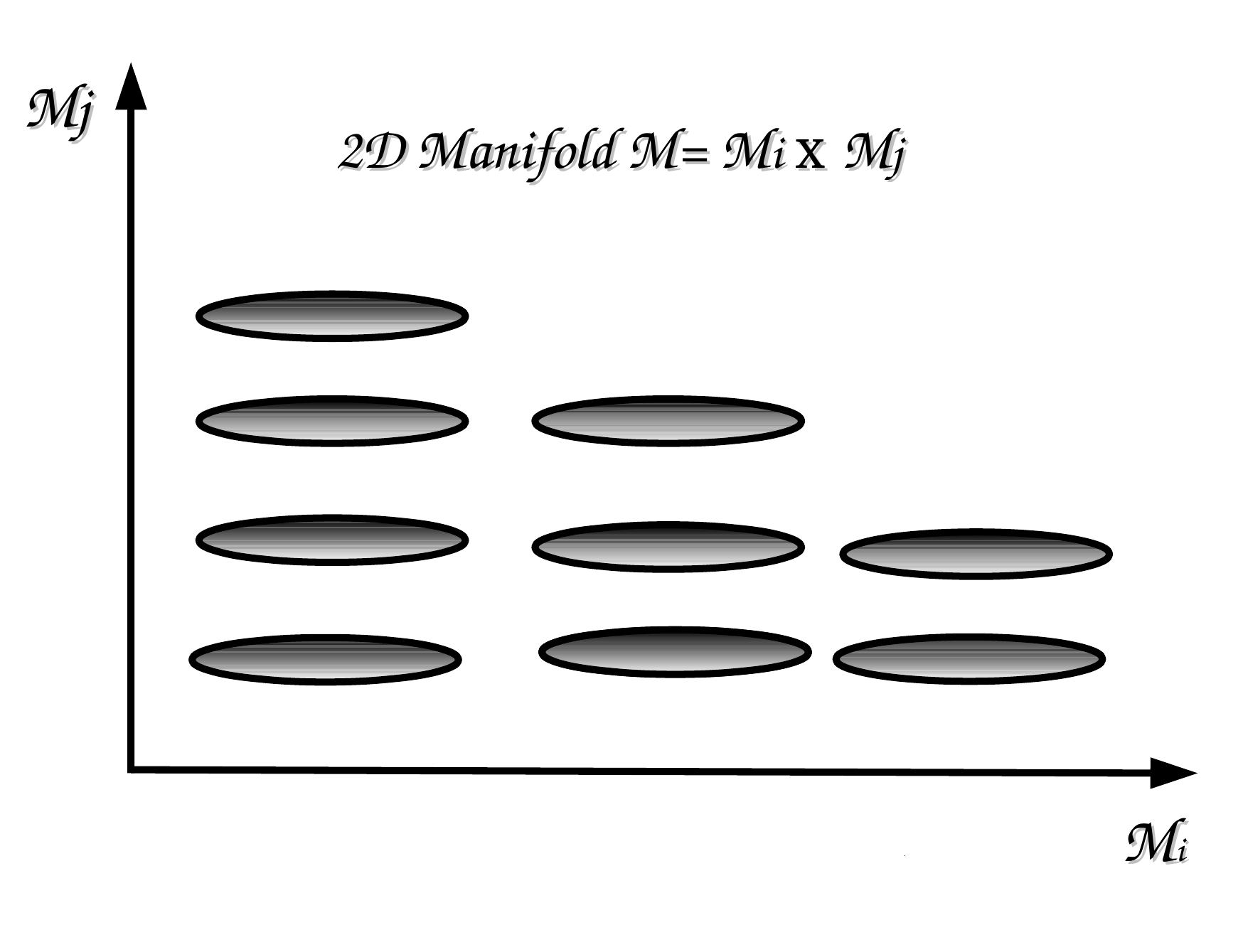}
   		\end{tabular}
   	\end{center}
   	\caption{  Constructing of 2-dimensional manifold by joining point like manifolds. }
   \end{figure*}

At fourth stage,  four point like manifolds or 2-dimensional manifolds join to each other and construct a 3-dimensional manifold (See Figure 7). Again, the potential of interactions between strings on the point like manifolds should be described by  delta functions. We get:

\begin{eqnarray}
	&& E_{M_{1}^{0}+M_{2}^{0}+M_{3}^{0}+M_{4}^{0}}=1=\nonumber\\&& \int_{M_{1}^{0}+M_{2}^{0}+M_{3}^{0}+M_{4}^{0}} d\tilde{X_{1}}^{I}d\tilde{X_{2}}_{I}d\tilde{X_{3}}_{I}d\tilde{X_{4}}^{I} \delta(\tilde{X_{1}}^{I})\delta(\tilde{X_{2}}^{I})\delta(\tilde{X_{3}}^{I})\delta(\tilde{X_{4}}^{I})=\nonumber\\&&  \int_{M_{1}^{0}+M_{2}^{0}+M_{3}^{0}+M_{4}^{0}} d\tilde{X_{1}}^{I}d\tilde{X_{2}}_{I} d\tilde{X_{3}}_{I}d\tilde{X_{4}}^{I}(\frac{1}{\sqrt{2\pi y_{1}}}e^{-\frac{\tilde{X_{1}}^{I}\tilde{X_{1}}_{I}}{2y_{1}}})(\frac{1}{\sqrt{2\pi y_{2}}}e^{-\frac{\tilde{X_{2}}^{I}\tilde{X_{2}}_{I}}{2y_{2}}})(\frac{1}{\sqrt{2\pi y_{3}}}e^{-\frac{\tilde{X_{3}}^{I}\tilde{X_{3}}_{I}}{2y_{3}}})(\frac{1}{\sqrt{2\pi y_{4}}}e^{-\frac{\tilde{X_{4}}^{I}\tilde{X_{4}}_{I}}{2y_{4}}})\nonumber\\&&	\label{spm18}
\end{eqnarray}

   It is clear that for 3-dimensional manifold, there are 3+1 integrations. Extra dimension is corresponded to time and totaly, we have 4-dimensional manifold. By using new difinition of  string fields  $\tilde{X}^{I}\longrightarrow \sqrt{2\pi y} X^{I}$, we obtain:

   \begin{eqnarray}
   	&& E_{M^{3}}=1= \int_{M_{1}^{0}+M_{2}^{0}+M_{3}^{0}+M_{4}^{0}} dX_{1}^{I}dX_{2 I}dX_{3 I} dX_{4}^{I}e^{-\pi X_{1}^{I}X_{1 I}}e^{-\pi X_{2}^{I}X_{2 I}}e^{-\pi X_{3}^{I}X_{3 I}}e^{-\pi X_{4}^{I}X_{4 I}}	\label{spm19}
   \end{eqnarray}

 When  point like manifolds join to each other and construct a 3-dimensional manifold,  strings will be  functions of coordinates of four manifolds ($X^{I}(y_{1},y_{2},y_{3},y_{4})$). In these conditions, equation (\ref{spm19}) can be taken the following shape:

  \begin{eqnarray}
  	&& E_{M^{3}}=1= \nonumber\\&& \int_{M_{1}^{0}+M_{2}^{0}+M_{3}^{0}+M_{4}^{0}}dy_{1}^{I}dy_{2 I}dy_{3 I}dy_{4}^{I} (\Sigma_{i,j,k,f=1,2,3,4}\frac{dX_{i}^{I}}{dy_{j}^{I}}\frac{dX_{j}^{I}}{dy_{k}^{I}}\frac{dX_{k}^{I}}{dy_{i}^{I}}\frac{dX_{i}^{I}}{dy_{f}^{I}})\times \nonumber\\&& e^{-\pi X_{1}^{I}X_{1 I}}e^{-\pi X_{2}^{I}X_{2 I}}e^{-\pi X_{3}^{I}X_{3 I}}e^{-\pi X_{4}^{I}X_{4 I}} +	\nonumber\\&& \int_{M_{1}^{0}+M_{2}^{0}+M_{3}^{0}+M_{4}^{0}}dy_{1}^{I}dy_{1 I}dy_{1 I}dy_{1}^{I} (\frac{dX_{1}^{I}}{dy_{1}^{I}}\frac{dX_{2 I}}{dy_{1 I}}\frac{dX_{3 I}}{dy_{1 I}}\frac{dX_{4 I}}{dy_{1 I}}) e^{-\pi X_{1}^{I}X_{1 I}}e^{-\pi X_{2}^{I}X_{2 I}}e^{-\pi X_{3}^{I}X_{3 I}}e^{-\pi X_{4}^{I}X_{4 I}}+\nonumber\\&&  \int_{M_{1}^{0}+M_{2}^{0}+M_{3}^{0}+M_{4}^{0}}dy_{2}^{I}dy_{2 I}dy_{2 I}dy_{2}^{I} (\frac{dX_{1}^{I}}{dy_{2}^{I}}\frac{dX_{2 I}}{dy_{2 I}}\frac{dX_{3 I}}{dy_{2 I}}\frac{dX_{4 I}}{dy_{2 I}}) e^{-\pi X_{1}^{I}X_{1 I}}e^{-\pi X_{2}^{I}X_{2 I}}e^{-\pi X_{3}^{I}X_{3 I}}e^{-\pi X_{4}^{I}X_{4 I}}+\nonumber\\&& \int_{M_{1}^{0}+M_{2}^{0}+M_{3}^{0}+M_{4}^{0}}dy_{3}^{I}dy_{3 I}dy_{3 I}dy_{3}^{I} (\frac{dX_{1}^{I}}{dy_{3}^{I}}\frac{dX_{2 I}}{dy_{3 I}}\frac{dX_{3 I}}{dy_{3 I}}\frac{dX_{4 I}}{dy_{3 I}}) e^{-\pi X_{1}^{I}X_{1 I}}e^{-\pi X_{2}^{I}X_{2 I}}e^{-\pi X_{3}^{I}X_{3 I}}e^{-\pi X_{4}^{I}X_{4 I}}+\nonumber\\&& \int_{M_{1}^{0}+M_{2}^{0}+M_{3}^{0}+M_{4}^{0}}dy_{4}^{I}dy_{4 I}dy_{4 I}dy_{4}^{I} (\frac{dX_{1}^{I}}{dy_{4}^{I}}\frac{dX_{2 I}}{dy_{4 I}}\frac{dX_{3 I}}{dy_{4 I}}\frac{dX_{4 I}}{dy_{4 I}}) e^{-\pi X_{1}^{I}X_{1 I}}e^{-\pi X_{2}^{I}X_{2 I}}e^{-\pi X_{3}^{I}X_{3 I}}e^{-\pi X_{4}^{I}X_{4 I}}\label{spm20}
  \end{eqnarray} 
   
  Similar to previous stages, using taylor mechanism and expanding exponential functions over the crossed points yields all terms of equation (\ref{spm7} and \ref{spm14}) in additional to following integration:

  \begin{eqnarray}
  	&& E_{M^{3}}=1= 3E_{M^{2}}-\nonumber\\&&  \int_{M_{1}^{0}+M_{2}^{0}+M_{3}^{0}+M_{4}^{0}}dy_{1}^{I}dy_{2 I}dy_{3 I}dy_{4}^{ I} (\Sigma_{i,j,k,f=1,2,3,4}\frac{dX_{i}^{I}}{dy_{j}^{I}}\frac{dX_{j}^{I}}{dy_{k}^{I}}\frac{dX_{k}^{I}}{dy_{i}^{I}}\frac{dX_{i}^{I}}{dy_{f}^{I}})\times \nonumber\\&& \pi^{4} (\Sigma_{i,j,k,f=1,2,4}(\frac{\partial}{\partial y_{i I}}(X_{i}^{I}X_{i I})+\frac{\partial}{\partial y_{i I}}\frac{\partial}{\partial y_{j}^{I}}(X_{i}^{I}X_{i I})+\frac{\partial}{\partial y_{i I}}\frac{\partial}{\partial y_{j}^{I}}\frac{\partial}{\partial y_{k I}}(X_{i}^{I}X_{i I})+\frac{\partial}{\partial y_{i I}}\frac{\partial}{\partial y_{j}^{I}}\frac{\partial}{\partial y_{k I}}\frac{\partial}{\partial y_{f}^{I}}(X_{i}^{I}X_{i I})+1)\times \nonumber\\&& (1+\frac{\partial}{\partial y_{j I}}(X_{j}^{I}X_{i I})+\frac{\partial}{\partial y_{i I}}\frac{\partial}{\partial y_{j}^{I}}(X_{j}^{I}X_{j I})+\frac{\partial}{\partial y_{i I}}\frac{\partial}{\partial y_{j}^{I}}\frac{\partial}{\partial y_{k I}}(X_{j}^{I}X_{j I})+ \frac{\partial}{\partial y_{j I}}\frac{\partial}{\partial y_{i}^{I}}\frac{\partial}{\partial y_{k I}}\frac{\partial}{\partial y_{f}^{I}}(X_{j}^{I}X_{j I}))\times \nonumber\\&& (\frac{\partial}{\partial y_{i I}}(X_{k}^{I}X_{k I})+\frac{\partial}{\partial y_{i I}}\frac{\partial}{\partial y_{j}^{I}}(X_{k}^{I}X_{k I})+\frac{\partial}{\partial y_{i I}}\frac{\partial}{\partial y_{j}^{I}}\frac{\partial}{\partial y_{k I}}(X_{k}^{I}X_{k I})+\frac{\partial}{\partial y_{k I}}\frac{\partial}{\partial y_{i}^{I}}\frac{\partial}{\partial y_{j I}}\frac{\partial}{\partial y_{f}^{I}}(X_{k}^{I}X_{k I})+1)\times\nonumber\\&&(\frac{\partial}{\partial y_{i I}}(X_{f}^{I}X_{f I})+\frac{\partial}{\partial y_{i I}}\frac{\partial}{\partial y_{j}^{I}}(X_{f}^{I}X_{f I})+\frac{\partial}{\partial y_{i I}}\frac{\partial}{\partial y_{j}^{I}}\frac{\partial}{\partial y_{k I}}(X_{f}^{I}X_{f I})+\frac{\partial}{\partial y_{f I}}\frac{\partial}{\partial y_{k I}}\frac{\partial}{\partial y_{i}^{I}}\frac{\partial}{\partial y_{j I}}(X_{f}^{I}X_{f I})+1)\times \nonumber\\&&\Big(\Sigma_{ijkf=0,1,2,3,4}(y_{4}^{I}-y_{0}^{I})^{f}(y_{3}^{I}-y_{0}^{I})^{i}(y_{2}^{I}-y_{0}^{I})^{j}(y_{1 I}-y_{0 I})^{k}\Big) +	...\label{spm21}
  \end{eqnarray} 
   
This equation indicates that  energy of 3-dimensional manifold can be derived by summing over energies of 2-dimensional manifolds plus extra energy which is needed for linking manifolds. Using definitions in equation (\ref{D4}, \ref{spm8} and \ref{spm15} ), we can obtain terms in above integration:

\begin{eqnarray}
&& X_{1}^{I}X_{1 I}=X_{2}^{I}X_{2 I}=X_{3}^{I}X_{3 I}= X_{4}^{I}X_{4 I} \nonumber\\&& \nonumber\\&& \nonumber\\&&     \Sigma_{ijkf}\frac{\partial}{\partial y_{f}^{I}}\frac{\partial}{\partial y_{k I}}\frac{\partial}{\partial y_{j I}}\frac{\partial}{\partial y_{i}^{I}}(X_{i}^{I}X_{i I})\approx \nonumber\\&& \frac{1}{2}\Sigma_{ijkf} (\epsilon_{I}^{ijkf}\frac{\partial}{\partial y_{f}^{I}}\Gamma_{ijk} +  \epsilon^{ijkf}\frac{\partial}{\partial y_{f}^{I}}\frac{\partial}{\partial y_{k I}}(F_{ij}))=\nonumber\\&& \frac{1}{2}\Sigma_{ijkf} (\epsilon_{I}^{ijkf}\frac{1}{2}[\frac{\partial}{\partial y_{f}^{I}}\Gamma_{ijk}-\frac{\partial}{\partial y_{i}^{I}}\Gamma_{fjk}]+\nonumber\\&& \epsilon_{I}\frac{1}{2}[\frac{\partial}{\partial y_{f}^{I}}\Gamma_{ijk}+\frac{\partial}{\partial y_{i}^{I}}\Gamma_{fjk}] + \epsilon^{ijkf} \frac{\partial}{\partial y_{f}^{I}}\frac{\partial}{\partial y_{k I}}(F_{ij}))\label{spm22}
\end{eqnarray}

 Again, we assume that point like manifolds are very closed to each other and thus, we can obtain:

\begin{eqnarray}
	&&  y_{4}^{I}-y_{0}^{I}= y_{3}^{I}-y_{0}^{I}= y_{2}^{I}-y_{0}^{I}=y_{1}^{I}-y_{0}^{I}=\sigma\longrightarrow \frac{1}{\pi}  \label{spmt16}
\end{eqnarray}

 Using equations ( \ref{spm7},\ref{spm8}, \ref{spm9}, \ref{spm10},\ref{spm14},\ref{spm15}, \ref{spm16}, \ref{spm17}, \ref{spm22}, \ref{spmt16} ) in equation (\ref{spm21}), we obtain:

\begin{eqnarray}
&& E_{M^{3}}=1= X^{I}X^{I}X_{I}X_{I}-\nonumber\\&&  \int_{M^{2}}d^{4}y\sqrt{-g}\Big(g^{ij}[\partial_{i}\Gamma^{k}_{jk}-\partial_{j}\Gamma^{k}_{ik} +\Gamma^{k}_{jf}\Gamma^{f}_{ik} -\Gamma^{k}_{if}\Gamma^{f}_{jk}]- \frac{1}{2}\partial_{i}\phi\partial^{i}\phi - \nonumber\\&&\frac{1}{4}\epsilon^{ijkm} F_{ij}F_{km}-\frac{1}{6}\epsilon^{ijkmln} F_{ij}F_{km}F_{ln}-\frac{1}{24}\epsilon^{ijkmlnfv} F_{ij}F_{km}F_{ln}F_{fv}-\nonumber\\&& \frac{1}{6}\epsilon^{ijkmln} \partial_{l}F_{ij}\partial_{n}F_{km}-\frac{1}{24}\epsilon^{ijkmlnfv} \partial_{l}\partial_{f}F_{ij}\partial_{n}\partial_{v}F_{km}+ V(\phi)+...  )= \nonumber\\&& X^{I}X^{I}X_{I}X_{I}-  \int_{M^{3}}d^{4}y\sqrt{-g}\Big(R-\frac{1}{2}\partial_{i}\phi\partial^{i}\phi - \frac{1}{4}\epsilon^{ijkm} F_{ij}F_{km}\nonumber\\&&-\frac{1}{6}\epsilon^{ijkmln} F_{ij}F_{km}F_{ln}-\frac{1}{24}\epsilon^{ijkmlnfv} F_{ij}F_{km}F_{ln}F_{fv}  - \frac{1}{2}g^{ij}g_{ij} -\nonumber\\&& \frac{1}{6}\epsilon^{ijkmln} \partial_{l}F_{ij}\partial_{n}F_{km}-\frac{1}{24}\epsilon^{ijkmlnfv} \partial_{l}\partial_{f}F_{ij}\partial_{n}\partial_{v}F_{km}+ V(\phi) +... ) \quad i,j=1,2,3,4\nonumber\\&& V(\phi)=1+ \frac{\phi^{2}}{2}-\frac{\phi^{3}}{6} + \frac{\phi^{4}}{24} \label{spm23}
\end{eqnarray} 

Above equation shows that by joining 2-dimensional manifolds and building 3-dimensional ones, gravity is emerged completely. In fact, energy of system devided to two parts, one is related to matters and usual quantum field theory and other is corresponded to curvatures which are direct signatures of gravity. Our universe is a three-dimensional manifold in space which has also one time dimension. For this reason, we can observe that energy of 3-dimensional manifold is the same of energy of matter fields in our universe.

\begin{figure*}[thbp]
	\begin{center}
		\begin{tabular}{rl}
			\includegraphics[width=8cm]{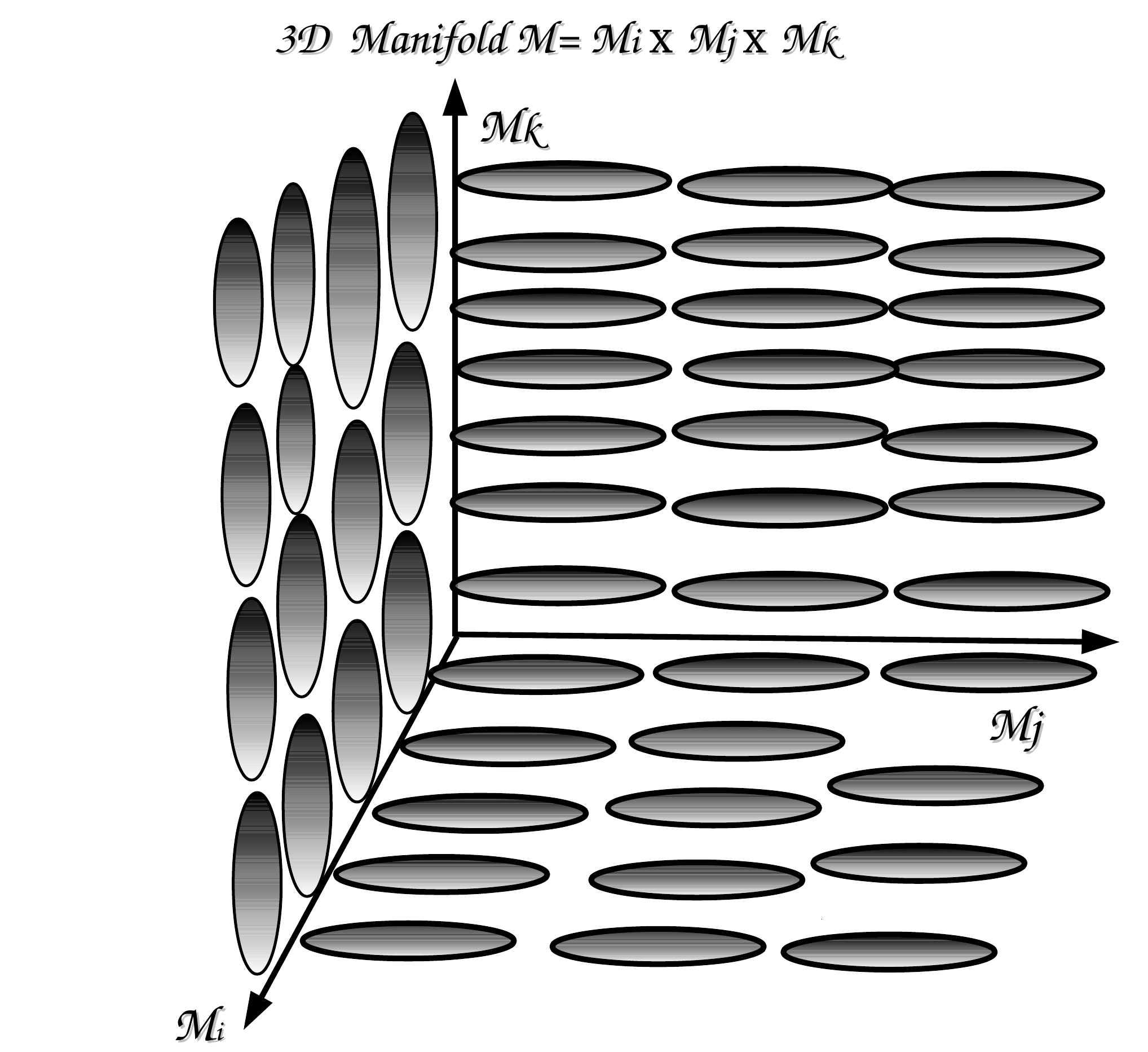}
		\end{tabular}
	\end{center}
	\caption{  Constructing of 3-dimensional manifold by joining point like manifolds. }
\end{figure*}

Now, we can generalize this mechanism to N-dimensional manifold in space (See Figure 8).  This manifold can be constructed by joining (N+1) point like manifolds, which the density potential of interactions between strings on them could be explained by  delta functions. In fact, all interactions on one point like manifold are concentrated on one point and thus total potential is approximately infinite. This potential which is zero in other points and infinite in one special point can be described by delta function. Integration over all these potentals, we obtain total energy of system:

\begin{eqnarray}
	&& E_{M_{i}^{0}}=\int_{M_{i}^{0}} d\tilde{X_{i}}^{I} V(\tilde{X_{i}}^{I})=\int_{M_{i}^{0}} d\tilde{X_{i}}^{I} \delta(\tilde{X_{i}}^{I})=1\Rightarrow \nonumber\\&&\nonumber\\&& \nonumber\\&& E_{M_{1}^{0}+...+M_{N+1}^{0}}=1=\nonumber\\&& \int_{M_{1}^{0}+...+M_{N+1}^{0}} d\tilde{X_{1}}^{I}...d\tilde{X_{N+1}}^{I} \delta(\tilde{X_{1}}^{I})...\delta(\tilde{X_{N+1}}^{I})=\nonumber\\&&  \int_{M_{1}^{0}+...+M_{N+1}^{0}} d\tilde{X_{1}}^{I}...d\tilde{X_{N+1}}^{I}(\frac{1}{\sqrt{2\pi y_{1}}}e^{-\frac{\tilde{X_{1}}^{I}\tilde{X_{1}}_{I}}{2y_{1}}})...(\frac{1}{\sqrt{2\pi y_{N+1}}}e^{-\frac{\tilde{X_{N+1}}^{I}\tilde{X_{N+1}}_{I}}{2y_{N+1}}})\nonumber\\&&	\label{spm24}
\end{eqnarray}
 
 In above equation, energy of manifold is normalized to one. It is observed that for N-dimensional manifold, there are N+1 integrations. Extra dimension is related to time and totaly, we have N+1-dimensional manifold. Applying new difinition of  string fields  $\tilde{X}^{I}\longrightarrow \sqrt{2\pi y} X^{I}$, we get:

 \begin{eqnarray}
 	&& E_{M^{N}}=1= \int_{M_{1}^{0}+....+M_{N+1}^{0}} dX_{1}^{I}... dX_{N+1}^{I}e^{-\pi X_{1}^{I}X_{1 I}}....e^{-\pi X_{N+1}^{I}X_{N+1 I}}	\label{spm25}
 \end{eqnarray} 
 
By joining  point like manifolds and constructing a N-dimensional manifold,  strings will be  functions of coordinates of N+1 manifolds ($X^{I}(y_{1}....y_{N+1})$). Thus, equation (\ref{spm25}) can be written the following form:

\begin{eqnarray}
	&& E_{M^{N}}=1= \nonumber\\&& \int_{M_{1}^{0}+...+M_{N+1}^{0}}dy_{1}^{I}...dy_{N+1}^{I} (\Sigma_{i_{1},...i_{N+1}=1...N+1}\frac{dX_{i_{2}}^{I}}{dy_{i_{1}}^{I}}\frac{dX_{i_{1}}^{I}}{dy_{i_{2}}^{I}}...\frac{dX_{i_{N+1}}^{I}}{dy_{i_{N}}^{I}}\frac{dX_{i_{N}}^{I}}{dy_{i_{N+1}}^{I}} e^{-\pi X_{i_{1}}^{I}X_{i_{1} I}}...e^{-\pi X_{i_{N+1}}^{I}X_{i_{N+1} I}} +	\nonumber\\&& \int_{M_{1}^{0}+...+M_{N+1}^{0}}dy_{i_{1}}^{I}...dy^{i_{1}I} (\frac{dX_{i_{1}}^{I}}{dy_{i_{1}}^{I}}....\frac{dX_{i_{N+1} I}}{dy_{i_{1} I}}) e^{-\pi X_{i_{1}}^{I}X_{i_{1} I}}...e^{-\pi X_{i_{N+1}}^{I}X_{i_{N+1} I}}+\nonumber\\&&...+\nonumber\\&&  \int_{M_{1}^{0}+...+M_{N+1}^{0}}dy_{i_{N+1}}^{I}...dy^{i_{N+1}I} (\frac{dX_{i_{1}}^{I}}{dy_{i_{N+1}}^{I}}....\frac{dX_{i_{N+1} I}}{dy_{i_{N+1} I}}) e^{-\pi X_{i_{1}}^{I}X_{i_{1} I}}...e^{-\pi X_{i_{N+1}}^{I}X_{i_{N+1} I}}\label{spm26}
\end{eqnarray} 

   Using taylor method and expanding exponential functions over the crossed points, we obtain all terms of equation (\ref{spm7}, \ref{spm14} and \ref{spm21}) in additional to following integration:

  \begin{eqnarray}
  	&& E_{M^{N}}=1= N E_{M^{N-1}}-\nonumber\\&& 
  	 \int_{M_{1}^{0}+...+M_{N+1}^{0}}dy_{1}^{I}...dy_{N+1}^{I} (\Sigma_{i_{1},...i_{N+1}=1...N+1}\frac{dX_{i_{2}}^{I}}{dy_{i_{1}}^{I}}\frac{dX_{i_{1}}^{I}}{dy_{i_{2}}^{I}}...\frac{dX_{i_{N+1}}^{I}}{dy_{i_{N}}^{I}}\frac{dX_{i_{N}}^{I}}{dy_{i_{N+1}}^{I}}\times \nonumber\\&& \pi^{N+1} (\Sigma_{i_{1},...,i_{N+1}=1,,,,N+1}(\frac{\partial}{\partial y_{i_{1} I}}(X_{i_{1}}^{I}X_{i_{1} I})+....+\frac{\partial}{\partial y_{i_{1} I}}...\frac{\partial}{\partial y_{i_{N+1}}^{I}}(X_{i_{1}}^{I}X_{i_{1} I})+1)\times \nonumber\\&&....\times \nonumber\\&&(\frac{\partial}{\partial y_{i_{1} I}}(X_{i_{N+1}}^{I}X_{i_{N+1} I})+...+\frac{\partial}{\partial y_{i_{1} I}}...\frac{\partial}{\partial y_{i_{N+1}}^{I}}(X_{i_{N+1}}^{I}X_{i_{N+1} I})+1)\times \nonumber\\&&\Big(\Sigma_{i_{1},...i_{N+1}=0...N+1}(y_{i_{N+1}}^{I}-y_{0}^{I})^{i_{N+1}}....(y_{1 I}-y_{0 I})^{i_{1}}\Big) +	...\label{spm27}
  \end{eqnarray}

This equation shows that  energy of N-dimensional manifold can be obtained by summing over energies of N-1-dimensional manifolds plus extra energy which is needed for linking manifolds. On the other hand, N-1-dimensional can be built from (N-2)-dimensional manifolds.  Also, this manifold can be produced by joining lower dimensional manifolds. Finally,  we can conlude that each manifold is constructed from zer0-dimensional or point-like manifolds. Extra energy which is created during producing manifolds can construct usual field theory in 4-dimensional universe.  Using definitions in equation (\ref{D4}, \ref{spm8}, \ref{spm15} and \ref{spm22} ), we can calculate terms in above integrations:

\begin{eqnarray}
	&& X_{1}^{I}X_{1 I}=....= X_{j}^{I}X_{j I} \nonumber\\&& \nonumber\\&& \nonumber\\&&     \Sigma_{i_{1},...i_{N+1}=1,..N+1}\frac{\partial}{\partial y_{i_{N+1}}^{I}}.....\frac{\partial}{\partial y_{i_{1}}^{I}}(X_{j}^{I}X_{j I})\approx \nonumber\\&& \frac{1}{2}\Sigma_{i_{1},...,i_{N+1}=1,..N+1}(\epsilon^{i_{1}...i_{N+1}}\frac{\partial}{\partial y_{i_{4}}^{I}}...\frac{\partial}{\partial y_{i_{N+1}}^{I}}(\Gamma_{i_{1}i_{2}i_{3}}) - \epsilon^{i_{1}...i_{N+1}}\frac{\partial}{\partial y_{i_{3}}^{I}}...\frac{\partial}{\partial y_{i_{N+1}}^{I}}(F_{i_{1}i_{2}}))+....\label{spm28}
\end{eqnarray} 

Again, we assume that point like manifolds are very closed to each other and thus, we can obtain:

\begin{eqnarray}
&&  y_{i}^{I}-y_{0}^{I}= ...y_{3}^{I}-y_{0}^{I}= y_{2}^{I}-y_{0}^{I}=y_{1}^{I}-y_{0}^{I}=\sigma\longrightarrow \frac{1}{\pi}  \label{spmtt16}
\end{eqnarray}

Using equations ( \ref{spm7},\ref{spm8}, \ref{spm9}, \ref{spm10},\ref{spm14},\ref{spm15}, \ref{spm16}, \ref{spm17},\ref{spm21}, \ref{spm22}, \ref{spmt16}, \ref{spm23},\ref{spm28},  \ref{spmtt16} ) in equation (\ref{spm27}), we obtain:

\begin{eqnarray}
&& E_{M^{N}}=1= (X^{I}X_{I})^{\frac{N}{2}}-\nonumber\\&&   \int_{M^{N}}d^{N+1}y\sqrt{-g}\Big(R-\frac{1}{2}\partial_{i}\phi\partial^{i}\phi - \frac{1}{4}\epsilon^{ijkm} F_{ij}F_{km}\nonumber\\&&-\frac{1}{6}\epsilon^{ijkmln} F_{ij}F_{km}F_{ln}-\frac{1}{24}\epsilon^{ijkmlnfv} F_{ij}F_{km}F_{ln}F_{fv}  - \frac{1}{2}g^{ij}g_{ij} -\nonumber\\&& \frac{1}{6}\epsilon^{ijkmln} \partial_{l}F_{ij}\partial_{n}F_{km}-\frac{1}{24}\epsilon^{ijkmlnfv} \partial_{l}\partial_{f}F_{ij}\partial_{n}\partial_{v}F_{km}-\nonumber\\&&.....-\nonumber\\&&\frac{1}{N(N-1)..1}\epsilon^{i_{1}i_{2}...i_{N}}( F_{i_{1}i_{2}}...F_{i_{N-3}i_{N-2}}F_{i_{N-1}i_{N}})...  + \frac{1}{N(N-1)..1}\epsilon^{i_{1}i_{2}...i_{N}} (R_{i_{1}i_{2}}...R_{i_{N-3}i_{N-2}}R_{i_{N-1}i_{N}})-\nonumber\\&&.....-\nonumber\\&&\frac{1}{N(N-1)..1}\epsilon^{i_{1}i_{2}...i_{N}}( \partial_{i_{1}}...\partial_{i_{N-2}}F_{i_{N-1}i_{N}})...  + \frac{1}{N(N-1)..1}\epsilon^{i_{1}i_{2}...i_{N}} (\partial_{i_{1}}...\partial_{i_{N-2}}R_{i_{N-1}i_{N}})+ V(\phi) +... \Big) \nonumber\\&& V(\phi)=1+ \frac{\phi^{2}}{2}..+\frac{\phi^{N}}{N(N-1)...1}  \label{spm29}
\end{eqnarray} 

This equation shows that by joining point like manifolds and constructing N-dimensional ones, the related energy in usual quantum field theory is appeared. Also, there are some extra terms related to higher order of gauge fields, curvatures and their derivatives that number of ordering depends on the number of dimension of manifold. This means that by increasing number of dimensions, order of fields and their derivatives grows.

\begin{figure*}[thbp]
	\begin{center}
		\begin{tabular}{rl}
			\includegraphics[width=8cm]{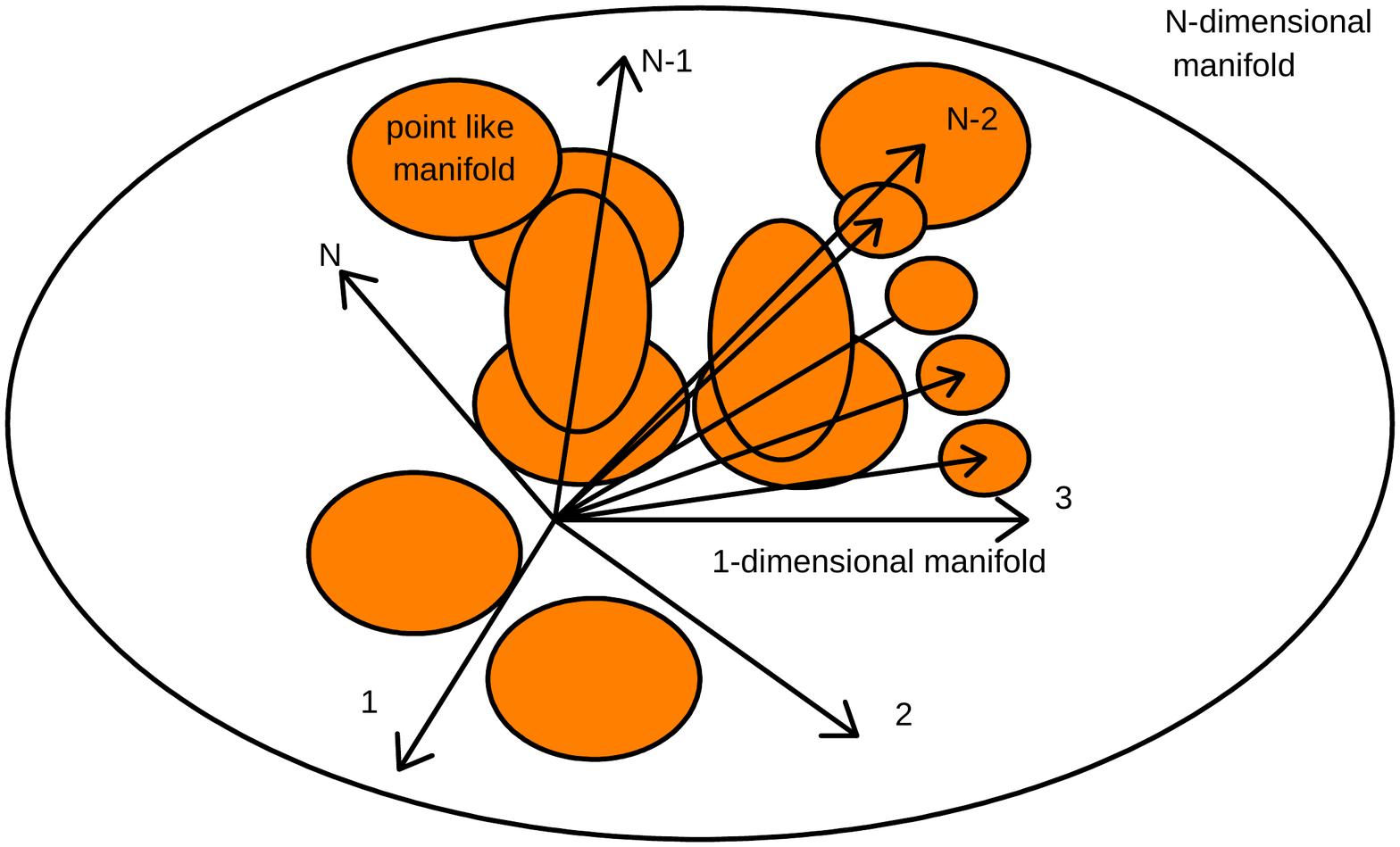}
		\end{tabular}
	\end{center}
	\caption{  Constructing of N-dimensional manifold by joining point like manifolds. }
\end{figure*}

Now, the question arises that what is the origin of the emergence of anomaly for some specian number of dimensions.  We will show that if N-dimensional manifold is broken to two or more manifolds, strings which are attached on each manifod depend on the coordinates of that manifold and miss the dependency on the coordinates of other manifolds. However, some strings are located between two manifolds and are functions of coordinates of both manifolds (See Figure 9). These strings produce the anomaly in each broken manifold which can be removed by regarding effect of other manifolds.

Using equation (\ref{spm27}) and regarding strings which are strengthed between two manifolds, we can obtain the energy of P-dimensional manifold:

\begin{eqnarray}
&& E_{M^{P},broken}=1= P E_{M^{P-1}}-\nonumber\\&& 
\int_{M_{1}^{0}+...+M_{P+1}^{0}}dy_{1}^{I}...dy_{P+1}^{I} (\Sigma_{i_{1},...i_{P+1}=1...P+1}\frac{dX_{i_{2}}^{I}}{dy_{i_{1}}^{I}}\frac{dX_{i_{1}}^{I}}{dy_{i_{2}}^{I}}...\frac{dX_{i_{P+1}}^{I}}{dy_{i_{P}}^{I}}\frac{dX_{i_{P}}^{I}}{dy_{i_{P+1}}^{I}}\times \nonumber\\&& \pi^{P+1} (\Sigma_{i_{1},...,i_{P+1}=1,,,,P+1}(\frac{\partial}{\partial y_{i_{1} I}}(X_{i_{1}}^{I}X_{i_{1} I})+....+\frac{\partial}{\partial y_{i_{1} I}}...\frac{\partial}{\partial y_{i_{P+1}}^{I}}(X_{i_{1}}^{I}X_{i_{1} I})+1)\times \nonumber\\&&....\times \nonumber\\&&(\frac{\partial}{\partial y_{i_{1} I}}(X_{i_{P+1}}^{I}X_{i_{P+1} I})+...+\frac{\partial}{\partial y_{i_{1} I}}...\frac{\partial}{\partial y_{i_{P+1}}^{I}}(X_{i_{P+1}}^{I}X_{i_{P+1} I})+1)\times \nonumber\\&& [\Sigma_{n=1}^{N-P} \Pi_{n} (\frac{\partial}{\partial y_{i_{1} I}}(X_{i_{P+1+n}}^{I}X_{i_{P+1+n} I})+...+\frac{\partial}{\partial y_{i_{1} I}}...\frac{\partial}{\partial y_{i_{P+1}}^{I}}(X_{i_{P+1+n}}^{I}X_{i_{P+1+n} I})+1)]\times \nonumber\\&&\Big(\Sigma_{i_{1},...i_{P+1}=0...P+1}(y_{i_{P+1}}^{I}-y_{0}^{I})^{i_{P+1}}....(y_{1 I}-y_{0 I})^{i_{1}}\Big) +	...\label{spm30}
\end{eqnarray}

In above equation, some extra strings have been aded to usual energy of manifold. These strings are produced during the breaking of parent manifold and the emergence of new child manifolds. They are functions of both coordinates of two child manifolds and stringthed between them.  We can rewrite the relation between energy of manifold which is produced by breaking of bigger manifold and the energy of manifold which is produced by joining point like manifolds:

\begin{eqnarray}
&&  X_{1}^{I}X_{1 I}=....= X_{j}^{I}X_{j I}\neq X_{i_{P+1+n}}^{I}X_{i_{P+1+n} I} \nonumber\\&&\nonumber\\&& \nonumber\\&& E_{M^{P},broken}= E_{M^{P}} \times \nonumber\\&& [\Sigma_{n=1}^{N-P} \Pi_{n} (\frac{\partial}{\partial y_{i_{1} I}}(X_{i_{P+1+n}}^{I}X_{i_{P+1+n} I})+...+\frac{\partial}{\partial y_{i_{1} I}}...\frac{\partial}{\partial y_{i_{P+1}}^{I}}(X_{i_{P+1+n}}^{I}X_{i_{P+1+n} I})+1)]\label{spm31}
\end{eqnarray}

Using equations ( \ref{spm7},\ref{spm8}, \ref{spm9}, \ref{spm10},\ref{spm14},\ref{spm15}, \ref{spm16}, \ref{spm17},\ref{spm21}, \ref{spm22}, \ref{spmt16}, \ref{spm23}, \ref{spm27}, \ref{spm28},  \ref{spmtt16},\ref{spm29} ) in equation (\ref{spm31}), we obtain:

\begin{eqnarray}
&& E_{M^{P}, broken}\approx\nonumber\\&&   -\int_{M^{P}}d^{P+1}y\sqrt{-g}\Big(R-\frac{1}{2}\partial_{i}\phi\partial^{i}\phi - \frac{1}{4}\epsilon^{ijkm} F_{ij}F_{km}\nonumber\\&&-\frac{1}{6}\epsilon^{ijkmln} F_{ij}F_{km}F_{ln}-\frac{1}{24}\epsilon^{ijkmlnfv} F_{ij}F_{km}F_{ln}F_{fv}  - \frac{1}{2}g^{ij}g_{ij} -\nonumber\\&& \frac{1}{6}\epsilon^{ijkmln} \partial_{l}F_{ij}\partial_{n}F_{km}-\frac{1}{24}\epsilon^{ijkmlnfv} \partial_{l}\partial_{f}F_{ij}\partial_{n}\partial_{v}F_{km}-\nonumber\\&&.....-\nonumber\\&&\frac{1}{P(P-1)..1}\epsilon^{i_{1}i_{2}...i_{p}}( F_{i_{1}i_{2}}...F_{i_{P-3}i_{P-2}}F_{i_{P-1}i_{P}})...  + \frac{1}{P(P-1)..1}\epsilon^{i_{1}i_{2}...i_{P}} (R_{i_{1}i_{2}}...R_{i_{P-3}i_{P-2}}R_{i_{P-1}i_{P}})-\nonumber\\&&.....-\nonumber\\&&\frac{1}{P(P-1)..1}\epsilon^{i_{1}i_{2}...i_{P}}( \partial_{i_{1}}...\partial_{i_{P-2}}F_{i_{P-1}i_{P}})...  + \frac{1}{P(P-1)..1}\epsilon^{i_{1}i_{2}...i_{P}} (\partial_{i_{1}}...\partial_{i_{P-2}}R_{i_{P-1}i_{P}})+... \Big)\times \nonumber\\&& \Big(1+ [\Sigma_{n=1}^{N-P} \Pi_{n} (\Sigma_{i_{1},,,i_{P}=1}^{P}(A_{i_{1}}^{i_{P+1+n}} +\partial_{i_{2}}\partial_{i_{3}}...\partial_{i_{P}}A_{i_{1}}^{i_{P+1+n}}+g_{i_{1}i_{P+1+n}}+\partial_{i_{2}}\partial_{i_{3}}...\partial_{i_{p}}g_{i_{1}i_{P+1+n}}))]\Big) \nonumber\\&&  \label{spm32}
\end{eqnarray} 

where $A_{i_{1}}^{i_{P+1+n}}$ and $g_{i_{1}i_{p+1+n}}$ play the role of gauge fields and graviton which move between manifolds. In fact, they are strings that strengthed between manifolds. Indice $i_{1}$ is related to one end of string  which is attached to the broken manifold and indice $i_{P+1+n}$ is corresponded to the end of string  which is located between manifolds.  We can show that these fields produce the anomaly. For this we calculate the gauge variation of energy in equation (\ref{spm32}):

\begin{eqnarray}
&&\delta_{A} E_{M^{P}, broken}\approx\nonumber\\&&   \int_{M^{P}}d^{P+1}y\sqrt{-g}\Big(\frac{1}{P(P-1)..1}\epsilon^{i_{1}i_{2}...i_{p}}( F_{i_{1}i_{2}}...F_{i_{P-3}i_{P-2}}F_{i_{P-1}i_{P}})  - \frac{1}{P(P-1)..1}\epsilon^{i_{1}i_{2}...i_{P}} (R_{i_{1}i_{2}}...R_{i_{P-3}i_{P-2}}R_{i_{P-1}i_{P}})+... \Big) \nonumber\\&& \delta_{A} E_{M^{N}}=0 \label{spm33}
\end{eqnarray} 

These results are in good agreement with the anomaly which has been obtained in Horava-Witten mechanism for 10-dimensional manifold in space plus one dimension in time \cite{b1,b2}. This means that the main reason for producing anomaly is the gauge fields and gravitons which fly between child manifolds. They are produced by breaking of parent manifolds and production of child manifolds. These fields can peoduce a new manifold which acts as a bridge between child manifolds (See Figure 10). For future goals, we  call this manifold as a Chern-Simons manifold.  In fact, if we sum over energies of child manifolds and Chern-Simons one, we obtain the energy of un-broken and initial N-dimensional Manifold which is anomaly free ($\delta_{A} E_{M^{N}}=0 $).

\begin{figure*}[thbp]
	
	\begin{center}
		\begin{tabular}{rl}
			\includegraphics[width=8cm]{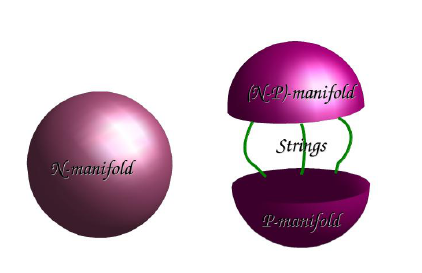}
		\end{tabular}
	\end{center}
	\caption{   N-dimensional manifold can be broken to two lower dimensional manifolds and some strings are attached to them. }
\end{figure*}

\begin{figure*}[thbp]
	
	\begin{center}
		\begin{tabular}{rl}
			\includegraphics[width=8cm]{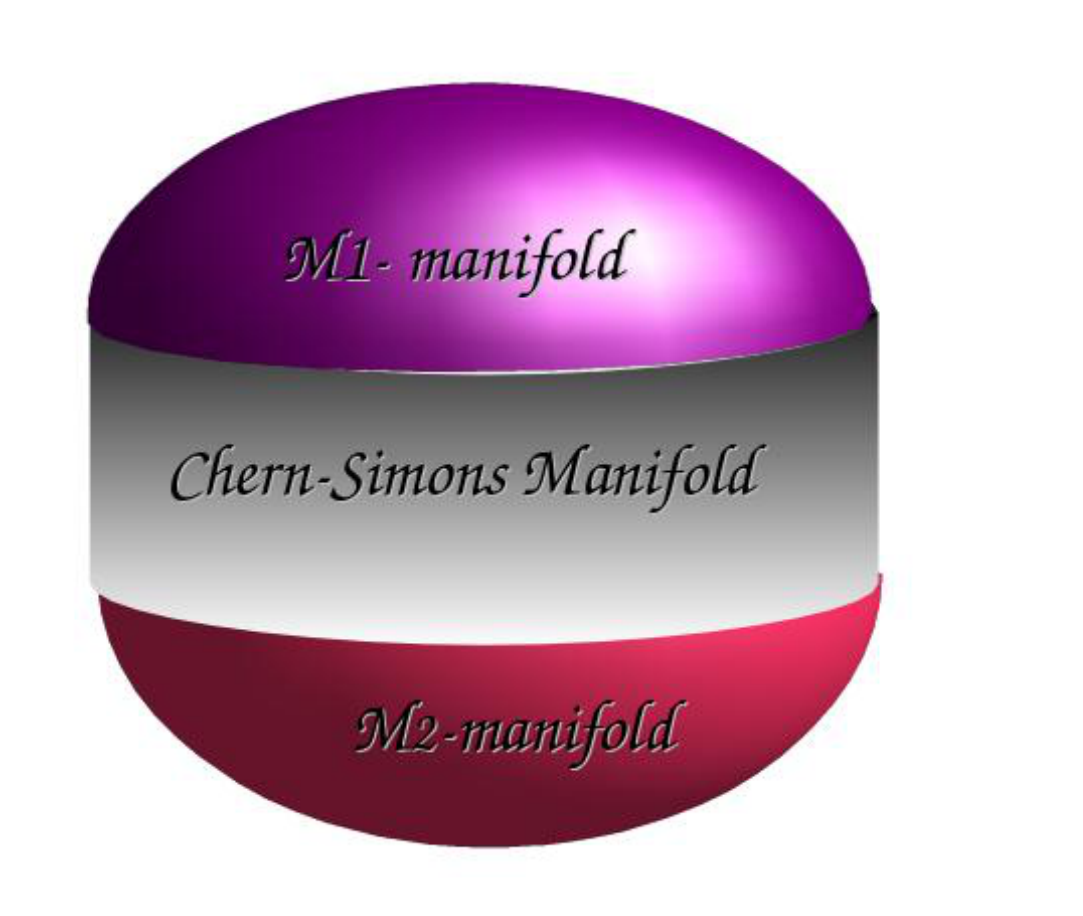}
		\end{tabular}
	\end{center}
	\caption{  Strings between child manifolds produce Chern-Simons manifold. }
\end{figure*}

Until now, we have discussed about the emergence of field strength with two indices during formation of N-dimensional manifolds. While in supergravity (See equation (\ref{s1})), we have G-fields with four indices. Now, the question arises that what is the origin of these fields? In equation (\ref{s2}), we observed that G-fields with four indices have a direct relation with F-terms with two indices. In fact, F-fields are produced by moving of two ends of strings. By joining two strings, an object with four ends is created which produces G-fields (See Figure 11).This means that number of indices or rank of field strengths is related to the number of ends of strings or objects which are produced by them. Using one of terms in  energy of P-dimensional manifold in equation (\ref{spm32}) and  the relation between F-terms and G-therms in equation (\ref{s2}), we can extract GG-terms in the action of supergravity in equation (\ref{s1}):

\begin{eqnarray}
&& G_{IJKL}\approx F_{[IJ}F_{KL]} \longrightarrow\nonumber\\&&\nonumber\\&&\nonumber\\&&  E_{M^{P}, broken}\approx  -\int_{M^{P}}d^{P+1}y\sqrt{-g}\Big(\frac{1}{24}\epsilon^{ijkmlnfv} F_{ij}F_{km}F_{ln}F_{fv}+.. ... \Big) \approx \nonumber\\&& \int_{M^{P}}d^{P+1}y\sqrt{-g}\Big(\frac{1}{24}\epsilon^{ijkmlnfv} G_{ijkm}G_{lnfv}+.. ... \Big)\label{spm34}
\end{eqnarray} 

This equation shows that by  joining F-fields, G-fields are emerged and produce GG-terms in the action of supergravity. In fact, by increasing dimension of manifolds, higher orders of F-fields are created that can cause to the production of G-terms with more indices.

\begin{figure*}[thbp]
	
	\begin{center}
		\begin{tabular}{rl}
			\includegraphics[width=8cm]{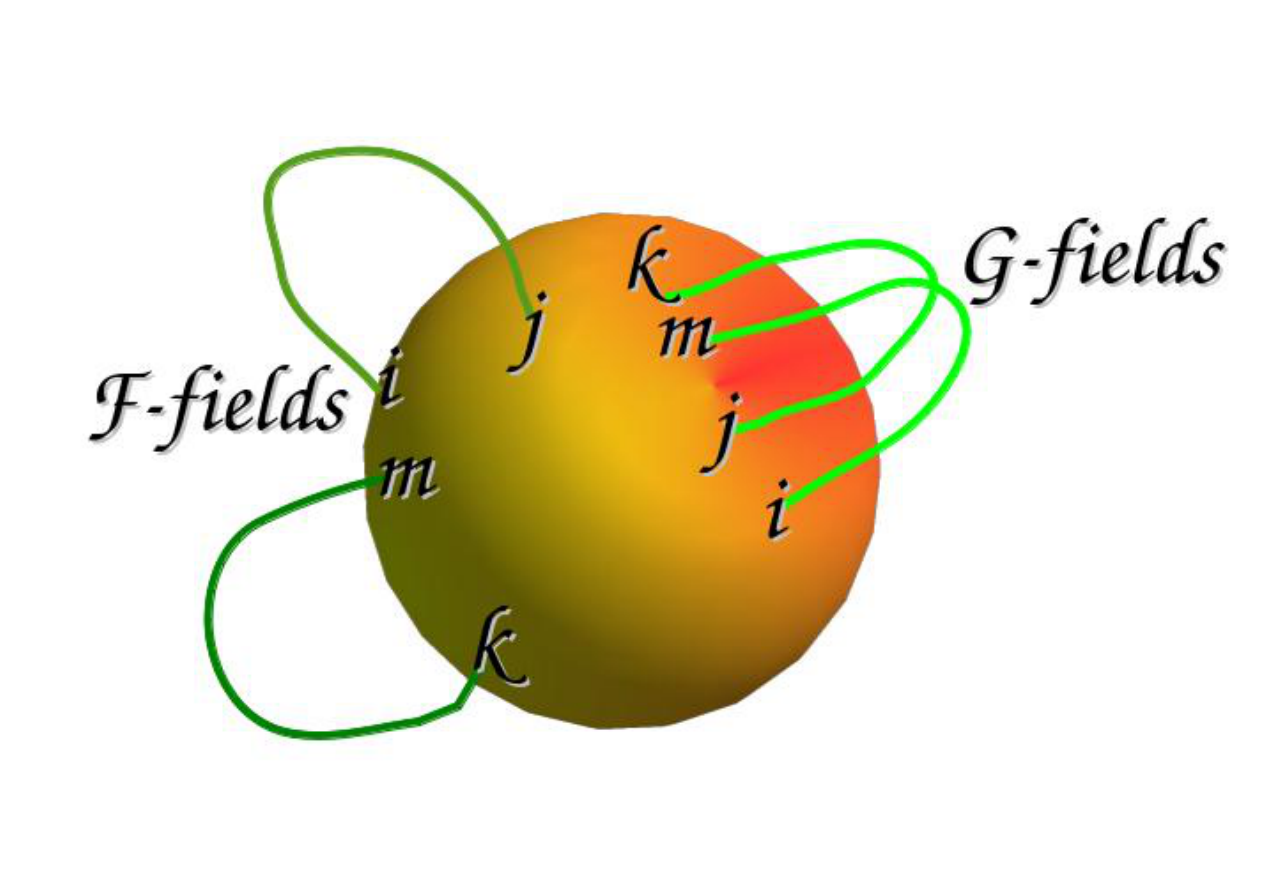}
		\end{tabular}
	\end{center}
	\caption{  Strings with two ends produce F-fields and objects with four indices, produce G-terms. }
\end{figure*}

 Now, we can show that by breaking manifolds, three ends of G-fields is located on P-dimensional manifold and one of ends of this field is placed on Chern-Simons manifold. In these conditions, C-fields in the action of supergravity (See equation (\ref{s1})) are created by breaking of G-fields (See Figure 12). Using one another terms of  energy of P-dimensional manifold in equation (\ref{spm32}) and  the relation between F-terms and C-therms in equation (\ref{s2}), we can extract CGG-terms in the action of supergravity in equation (\ref{s1}):

  \begin{eqnarray}
  && C_{i_{2}i_{3}i_{1}}^{i_{P+1+n}}=\epsilon^{i_{2}i_{3}i_{1}} F_{i_{2}i_{3}}A_{i_{1}}^{i_{P+1+n}}\longrightarrow\nonumber\\&&\nonumber\\&&\nonumber\\&& E_{CGG..G}\approx   -\int_{M^{P}}d^{P+1}y\sqrt{-g}\Big(\frac{1}{P(P-1)..1}\epsilon^{i_{1}i_{2}...i_{p}}( F_{i_{1}i_{2}}...F_{i_{P-3}i_{P-2}}F_{i_{P-1}i_{P}})+...  \Big)\times \nonumber\\&& \Big(1+ [\Sigma_{n=1}^{N-P} \Pi_{n} (\Sigma_{i_{1},,,i_{P}=1}^{P}(A_{i_{1}}^{i_{P+1+n}} +\partial_{i_{2}}\partial_{i_{3}}...\partial_{i_{P}}A_{i_{1}}^{i_{P+1+n}}+g_{i_{1}i_{P+1+n}}+\partial_{i_{2}}\partial_{i_{3}}...\partial_{i_{p}}g_{i_{1}i_{P+1+n}}))]\Big)\approx \nonumber\\&& -\int_{M^{P}}d^{P+1}y\sqrt{-g}\Big(\frac{1}{P(P-1)..1}\epsilon^{i_{1}i_{2}...i_{p}}( F_{i_{1}i_{2}}...F_{i_{P-3}i_{P-2}}F_{i_{P-1}i_{P}})+...  \Big)A_{i_{1}}^{i_{P+1+n}}+.. \approx\nonumber\\&& -\int_{M^{P}}d^{P+1}y\sqrt{-g}\Big(\frac{1}{P(P-1)..1}\epsilon^{i_{1}i_{2}...i_{p}}(C_{i_{1}i_{2}i_{3}}^{i_{P+1+n}} G_{i_{4}i_{5}i_{6}i_{7}}..G_{i_{P-2}i_{P-1}i_{P}i_{P+1}}+...  \Big)  \label{spm35}
  \end{eqnarray} 
  
 For P=10, above energy can be reduced to CGG-terms in eleven-dimensional supergravity:

 \begin{eqnarray}
&& E_{CGG} \approx -\int_{M^{11}}d^{11}y\sqrt{-g}\Big(\frac{1}{10(9)..1}\epsilon^{i_{1}i_{2}...i_{11}}(C_{i_{1}i_{2}i_{3}}^{i_{11+n}} G_{i_{4}i_{5}i_{6}i_{7}}G_{i_{8}i_{9}i_{10}i_{11}}+...  \Big)  \label{spm36}
 \end{eqnarray} 
 
Above results show that  by breaking manifolds, one of ends of some of G-fields are placed on Chern-Simons manifold. These fields are observed only with three inices from an observer on P-dimensional manifold. These types of fields are known as  C-fields. Then, these fields interact with other G-fields on the manifold and produce CGG terms in the action of supergravity. Extra indice $i_{P+1+n}$ on C-fields indicates that one of ends of this field is located outside of broken manifold and on Chern-Simons manifold (See Figure 12).

\begin{figure*}[thbp]
	
	\begin{center}
		\begin{tabular}{rl}
			\includegraphics[width=8cm]{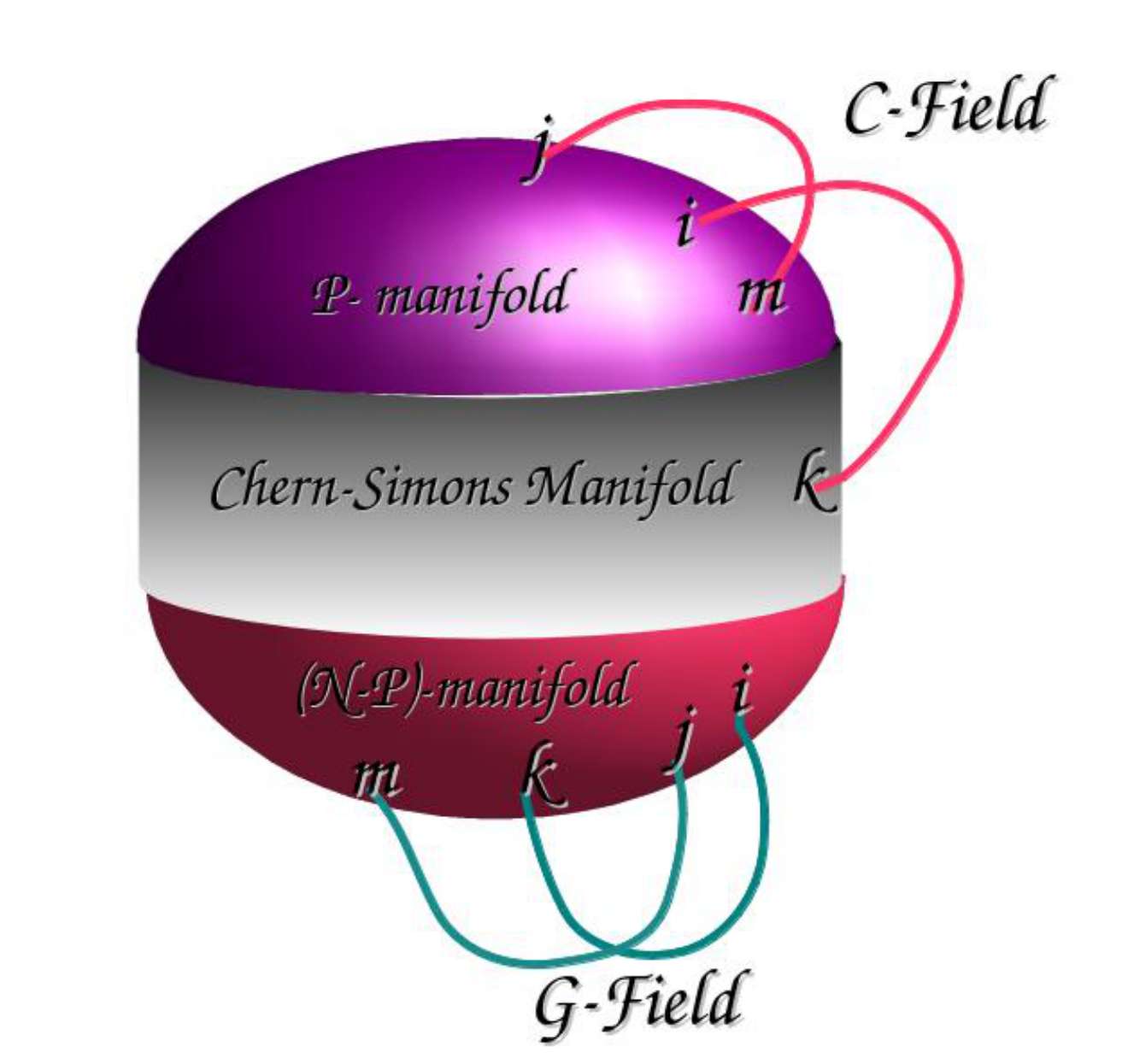}
		\end{tabular}
	\end{center}
	\caption{  C-fields have three ends on P-dimensional manifold and one end on Chern-Simons manifold. }
\end{figure*}

      Until now, we have have obtained the explicit relation between scalars, gauge fields, gravitons, G-fields, C-fields and strings. We also calculated the energy of gravity and matter fields which live on N-dimensional manifold. We have discussed that by breaking N-dimensional manifold, lower dimensional manifolds plus Chern-Simons manifold are created. We have shown that strings which are strengthed between two manifolds are the main cause of the emergence of anomaly.  Now, we want to limit our selves to two 11-dimensional manifolds in Horava-Witten mechanism plus extra three dimensional Chern-Simons manifold which connect them.  We obtain the exact terms of MOG gravity and show that this type of modified gravity is anomaly free.
      
      Using equation (\ref{spm7},\ref{spm8}, \ref{spm9}, \ref{spm10},\ref{spm14},\ref{spm15}, \ref{spm16}, \ref{spm17},\ref{spm21}, \ref{spm22}, \ref{spmt16}, \ref{spm23}, \ref{spm27}, \ref{spm28},  \ref{spmtt16}, \ref{spm29}) for strings, we can obtain the relation between fields and strings \cite{b9,A1}:

      \begin{eqnarray}
      &&\frac{1}{E_{M^{N}}} =E_{M^{N}}=1 \Rightarrow\nonumber\\&& [ (X^{I}X_{I})^{\frac{N}{2}}-   \int_{M^{N}}d^{N+1}y\sqrt{-g}\Big(R-\frac{1}{2}\partial_{i}\phi\partial^{i}\phi - \frac{1}{4}\epsilon^{ijkm} F_{ij}F_{km}\nonumber\\&&-\frac{1}{6}\epsilon^{ijkmln} F_{ij}F_{km}F_{ln}-\frac{1}{24}\epsilon^{ijkmlnfv} F_{ij}F_{km}F_{ln}F_{fv}  - \frac{1}{2}g^{ij}g_{ij} -\nonumber\\&& \frac{1}{6}\epsilon^{ijkmln} \partial_{l}F_{ij}\partial_{n}F_{km}-\frac{1}{24}\epsilon^{ijkmlnfv} \partial_{l}\partial_{f}F_{ij}\partial_{n}\partial_{v}F_{km}-\nonumber\\&&.....-\nonumber\\&&\frac{1}{N(N-1)..1}\epsilon^{i_{1}i_{2}...i_{N}}( F_{i_{1}i_{2}}...F_{i_{N-3}i_{N-2}}F_{i_{N-1}i_{N}})...  + \frac{1}{N(N-1)..1}\epsilon^{i_{1}i_{2}...i_{N}} (R_{i_{1}i_{2}}...R_{i_{N-3}i_{N-2}}R_{i_{N-1}i_{N}})-\nonumber\\&&.....-\nonumber\\&&\frac{1}{N(N-1)..1}\epsilon^{i_{1}i_{2}...i_{N}}( \partial_{i_{1}}...\partial_{i_{N-2}}F_{i_{N-1}i_{N}})...  + \frac{1}{N(N-1)..1}\epsilon^{i_{1}i_{2}...i_{N}} (\partial_{i_{1}}...\partial_{i_{N-2}}R_{i_{N-1}i_{N}})+ V(\phi) +... \Big)]^{-1}= \nonumber\\&& [ (X^{I}X_{I})^{\frac{N}{2}}-   \int_{M^{N}}d^{N+1}y\sqrt{-g}\Big(R-\frac{1}{2}\partial_{i}\phi\partial^{i}\phi - \frac{1}{4}\epsilon^{ijkm} F_{ij}F_{km}\nonumber\\&&-\frac{1}{6}\epsilon^{ijkmln} F_{ij}F_{km}F_{ln}-\frac{1}{24}\epsilon^{ijkmlnfv} F_{ij}F_{km}F_{ln}F_{fv}  - \frac{1}{2}g^{ij}g_{ij} -\nonumber\\&& \frac{1}{6}\epsilon^{ijkmln} \partial_{l}F_{ij}\partial_{n}F_{km}-\frac{1}{24}\epsilon^{ijkmlnfv} \partial_{l}\partial_{f}F_{ij}\partial_{n}\partial_{v}F_{km}-\nonumber\\&&.....-\nonumber\\&&\frac{1}{N(N-1)..1}\epsilon^{i_{1}i_{2}...i_{N}}( F_{i_{1}i_{2}}...F_{i_{N-3}i_{N-2}}F_{i_{N-1}i_{N}})...  + \frac{1}{N(N-1)..1}\epsilon^{i_{1}i_{2}...i_{N}} (R_{i_{1}i_{2}}...R_{i_{N-3}i_{N-2}}R_{i_{N-1}i_{N}})-\nonumber\\&&.....-\nonumber\\&&\frac{1}{N(N-1)..1}\epsilon^{i_{1}i_{2}...i_{N}}( \partial_{i_{1}}...\partial_{i_{N-2}}F_{i_{N-1}i_{N}})...  + \frac{1}{N(N-1)..1}\epsilon^{i_{1}i_{2}...i_{N}} (\partial_{i_{1}}...\partial_{i_{N-2}}R_{i_{N-1}i_{N}})+ V(\phi) +... \Big)]
      \label{sye4}
  \end{eqnarray}

Assuming that fields change slowly respect to coordinates and using of $ \hat{G}_{IJ}=R_{IJ}-\frac{1}{2}R g_{IJ}$, we can solve above equation and obtain the better result respect to equation (\ref{D4}) for relation between strings and fields:
      
      \begin{eqnarray}
      && I^{J}=\epsilon^{J}I  \nonumber\\ &&
            X^{I}= I^{I} + y^{I}+ \epsilon^{I}\phi + A^{I} + y_{J}( F^{IJ}- R^{IJ}+\partial^{I}\phi\partial^{J}\phi-\epsilon^{IJ}\Sigma_{n}\hat{G}^{-\frac{n}{2}-1}+...)
\label{s4}
\end{eqnarray}

where $\phi$ is the scalar field, and  $A^{I}$ is the gauge field and $\Gamma$ has the relation with the curvature (R)
and $I^J$ is a unit vector in direction of (J-th) dimension.  Previously, we have shown that before joining point like manifolds and formation of higher dimensional manifolds, the origin of all matter and strings are the same and they are equal to unit vectors ($I^{J}=I\epsilon^{J}$ ($\epsilon^{J}$ is a symbole that caries indice of dimension)).  Then, by  constructing N-dimensional manifolds by joining point like manifolds, some extra fields ($\phi,A^{I}$) are appeared and interact with each other.  Before joining point like manifolds, we have a symmetry which can be explained in terms of unit vectors. In the static state, all strings can be described by unit vectors. When manifolds interact with each other, the symmetry is broken and fields emerge. 

By above definition, we can show that gauge fields and curvatures in energy of manifolds like equation (\ref{spm29}) can be replaced by Poisson brackets. These brackets help us to propose a unified model for writing interacting terms in action or energy.  In fact, using of Poisson brackets instead of many various fields in action, make the complicated calculations very easy. For this, using equation (\ref{s4}), we obtain following relation:

 \begin{eqnarray}
       \{ X^{I},X^J\}&=&\Sigma_{I,J}\varepsilon^{I'J'}\frac{\partial X^{I}}{\partial y^{I'}}\frac{\partial X^{J}}{\partial y^{J'}}\nonumber\\
       &=&F^{IJ}- R^{IJ}+\partial^{I}\phi\partial^{J}\phi-\epsilon^{IJ}\Sigma_{n}\hat{G}^{-\frac{n}{2}-1}+....\label{syee4}
       \end{eqnarray}
where $\epsilon$ is a symbole which carries indices of dimensions.  Above equation shows that various fields like gauge fields and curvatures in energy or action can be replaced by a unified shape of poisson brackets.  Using this technique makes the calculations easy and help us to find ways for removing the anomaly from manifolds. Using 4-dimensional instead of 2-dimensional brackets, we can obtain the GG term $G_{IJKL}G^{IJKL}$ in supergravity in terms of strings ($X$):

          \begin{eqnarray}
          G^{IJKL}&=& \{ X^{I},X^{J},X^{K},X^{L} \} =
          \varepsilon^{I'J'K'L'}\frac{\partial X^{I}}{\partial y^{I'}}\frac{\partial X^{J}}{\partial y^{J'}}\frac{\partial X^{K}}{\partial y^{K'}}\frac{\partial X^{L}}{\partial y^{L'}}\nonumber\\
          &\Downarrow&\nonumber\\
          \int d^{11}x\sqrt{g}\Big(G_{IJKL}G^{IJKL}\Big)&=& 
          \int d^{11}x\sqrt{g}\Big(
          \varepsilon_{I'J'K'L'}\frac{\partial X_{I}}{\partial y_{I'}}\frac{\partial X_{J}}{\partial y_{J'}}\frac{\partial X_{K}}{\partial y_{K'}}\frac{\partial X_{L}}{\partial y_{L'}}
          \varepsilon^{I''J''K''L''}\frac{\partial X^{I}}{\partial y^{I''}}\frac{\partial X^{J}}{\partial y^{J''}}\frac{\partial X^{K}}{\partial y^{K''}}\frac{\partial X^{L}}{\partial y^{L''}}\Big).\label{s13}
          \end{eqnarray}

       Equation (\ref{s13}) helps us extract the CGG terms from the GG terms in supergravity. To this end, we will add a 3-dimensional manifold  related to Lie-three-algebra to the 11-dimensional supergravity by using the properties of strings ($X$) in Nambu-Poisson brackets \cite{A1}:

           \begin{eqnarray}
           X^{I}&=&  I^{I} + y^{I}+ \epsilon^{I}\phi + A^{I} + y_{J}( F^{IJ}- R^{IJ}+\partial^{I}\phi\partial^{J}\phi-\epsilon^{IJ}\Sigma_{n}\hat{G}^{-\frac{n}{2}-1}+...)\nonumber\\
           &\Downarrow&\nonumber\\
           \frac{\partial X^{I_{5}}}{\partial y^{I_{5}}}&\approx&\delta ( y^{I_{5}})+... \quad \frac{\partial X^{I_{6}}}{\partial y^{I_{6}}}\approx\delta ( y^{I_{6}})+... \quad \frac{\partial X^{I_{7}}}{\partial y^{I_{7}}}\approx\delta ( y^{I_{7}})+...,\\
           \int_{M^{N=3}}&\rightarrow&\int_{y^{I_{5}}+y^{I_{6}}+y^{I_{7}}}\int dy^{I_{5}}dy^{I_{6}}dy^{I_{7}}\varepsilon^{I'_{5}I'_{6}I'_{7}}\frac{\partial X^{I_{5}}}{\partial y^{I'_{5}}}\frac{\partial X^{I_{6}}}{\partial y^{I'_{6}}}\frac{\partial X^{I_{7}}}{\partial y^{I'_{7}}}=1+..., \label{s14}
           \end{eqnarray}
where ellipses (...) were used to represent higher-order derivatives. The integration is over a 3-dimensional manifold with coordinates ($y^{I_{5}},y^{I_{6}},y^{I_{7}}$) and consequently, the integration can be shown by $\int_{y^{I_{5}}+y^{I_{6}}+y^{I_{7}}}=\int dy^{I_{5}}\int dy^{I_{6}}\int dy^{I_{7}}$). This result shows that ignoring fluctuations of space that cause the production of fields, the result of integration over each 3-dimensional manifold tends to one. When, we add one manifold to another, the integration will be the product of integration over each manifold.

Extending the manifold over additional dimensions extends the integration volume element.  By extending the 11-dimensional manifold in Eq.~(\ref{s13}) with the 3-dimensional manifold of Eq.~(\ref{s14}), we get

            \begin{eqnarray}
            && \int_{M^{11}}d^{11}x\sqrt{g}\Big(G_{I_{1}I_{2}I_{3}I_{4}}G^{I_{1}I_{2}I_{3}I_{4}}\Big)\times \int_{M^{N=3}}\nonumber\\
            &=&
            \int_{M^{11}+y^{I_{5}}+y^{I_{6}}+y^{I_{7}}}\int d^{11}x\int dy^{I_{5}}dy^{I_{6}}dy^{I_{7}}\sqrt{g}\epsilon^{I'_{5}I'_{6}I'_{7}} G_{I_{1}I_{2}I_{3}I_{4}}G^{I_{1}I_{2}I_{3}I_{4}}\frac{\partial X^{I_{5}}}{\partial y^{I'_{5}}}\frac{\partial X^{I_{6}}}{\partial y^{I'_{6}}}\frac{\partial X^{I_{7}}}{\partial y^{I'_{7}}},
\end{eqnarray}
where we notice that after making the identification
            \begin{eqnarray}
            C^{I_{5}I_{6}I_{7}}&=& 
            \epsilon^{I'_{5}I'_{6}I'_{7}}\frac{\partial X^{I_{5}}}{\partial y^{I'_{5}}}\frac{\partial X^{I_{6}}}{\partial y^{I'_{6}}}\frac{\partial X^{I_{7}}}{\partial y^{I'_{7}}},\label{s15}
            \end{eqnarray}
we recover the CGG action in $11+3$ dimensions:
\begin{equation}
S^{N=11+3}_{\rm CGG}=\int_{M^{11}+y^{I_{5}}+y^{I_{6}}+y^{I_{7}}}\int d^{11}x\int dy^{I_{5}}dy^{I_{6}}dy^{I_{7}}\sqrt{g} G_{I_{1}I_{2}I_{3}I_{4}}G^{I_{1}I_{2}I_{3}I_{4}}
C^{I_{5}I_{6}I_{7}}.\label{sop15}
\end{equation}

This equation has three interesting results:
\begin{inparaenum}[1.]
\item The CGG term appears in the supergravity action as a result of adding a 3-dimensional manifold, 
\item Combining the 11-dimensional supergravity with the 3-dimensional manifold yields 14-dimensional supergravity.
\item The shape of C-term is now clear in terms of string fields ($X^{I}$).
\end{inparaenum}

       To verify that  the theory is correct, we should be able to get back the gauge variation of the CGG-action in Eq.~(\ref{s3}) in terms of fields strengths and curvature. To this end, using Eqs.~(\ref{s14}) and (\ref{s15}), we can calculate the gauge variation of C \cite{A1}:

             \begin{eqnarray}
             X^{I}&=&I^{I} + y^{I} + \epsilon^{I}\phi + A^{I} + y_{J}( F^{IJ}- R^{IJ}+\partial^{I}\phi\partial^{J}\phi-\epsilon^{IJ}\Sigma_{n}\hat{G}^{-\frac{n}{2}-1}+...)\nonumber\\
             &\Downarrow&\nonumber\\
             \frac{\partial \delta_{A} X^{I}}{\partial y^{I}}&=&\delta ( y^{I}) \nonumber\\
             &\Downarrow&\nonumber\\
             \int_{M^{N=2}+M^{11}}\int d^{11}x\int dy^{I_{5}}dy^{I_{6}}dy^{I_{7}}\delta_{A} C^{I_{5}I_{6}I_{7}} &=&\int_{M^{N=3}+M^{11}}\int d^{11}x\int dy^{I_{5}}dy^{I_{6}} dy^{I_{7}} \Sigma_{I'_{5}I'_{6}I'_{7}}\varepsilon^{I'_{5}I'_{6}I'_{7}}\delta_{A}(\frac{\partial X^{I_{5}}}{\partial y^{I'_{5}}}\frac{\partial X^{I_{6}}}{\partial y^{I'_{6}}}\frac{\partial X^{I_{7}}}{\partial y^{I'_{7}}})\nonumber\\
             &=& \int_{M^{N=2}+M^{11}}\int d^{11}x\int dy^{I_{5}}dy^{I_{6}}\Sigma_{I'_{5}I'_{6}}\varepsilon^{I'_{5}I'_{6}}(\frac{\partial X^{I_{5}}}{\partial y^{I'_{5}}}\frac{\partial X^{I_{6}}}{\partial y^{I'_{6}}})\nonumber\\
             &=&\int_{M^{N=2}+M^{11}}\int d^{11}x\int dy^{I}dy^{J}(F^{IJ}- R^{IJ}\nonumber\\&+&\partial^{I}\phi\partial^{J}\phi-\varepsilon^{IJ}\Sigma_{n=1}^{\infty}\hat{G}^{-\frac{n}{2}-1})+...\nonumber\\
             &=&\int_{M^{N=2}+M^{11}}\int d^{11}x\int dy^{I}dy^{J}(F^{IJ}- R^{IJ} \nonumber\\&+&\partial^{I}\phi\partial^{J}\phi-\varepsilon^{IJ}\hat{G}^{-1} +...).\label{s16}
             \end{eqnarray}
where the ellipses (...) represent higher-order derivatives with respect to fields. 
Using Eqs.~(\ref{s16}) and (\ref{s13})  we can calculate the gauge variation of the CGG action in equation of (\ref{sop15}):
\begin{eqnarray}
\delta S^{N=11+3}_{\rm CGG}&=&
\delta\int_{M^{11}+M^{N=3}}d^{14}x\sqrt{g}\epsilon_{I_{1}I_{2}I_{3}I_{4}I'_{1}I'_{2}I'_{3}I'_{4}I_{5}I_{6}I_{7}}\epsilon^{\tilde{I}_{5}\tilde{I}_{6}\tilde{I}_{7}}(\frac{\partial X^{I_{5}}}{\partial y^{\tilde{I}_{5}}}\frac{\partial X^{I_{6}}}{\partial y^{\tilde{I}_{6}}}\frac{\partial X^{I_{7}}}{\partial y^{\tilde{I}_{7}}}) G^{I_{1}I_{2}I_{3}I_{4}}G^{I'_{1}I'_{2}I'_{3}I'_{4}}\nonumber\\
&=&\delta\int_{M^{11}+M^{N=3}}d^{14}x\sqrt{g}\epsilon_{I_{1}I_{2}I_{3}I_{4}I'_{1}I'_{2}I'_{3}I'_{4}I_{5}I_{6}I_{7}}\epsilon^{\tilde{I}_{5}\tilde{I}_{6}\tilde{I}_{7}}(\frac{\partial X^{I_{5}}}{\partial y^{\tilde{I}_{5}}}\frac{\partial X^{I_{6}}}{\partial y^{\tilde{I}_{6}}}\frac{\partial X^{I_{7}}}{\partial y^{\tilde{I}_{7}}})\nonumber\\
&&\hskip 0.5in{}\times(\epsilon^{\tilde{I}_{1}\tilde{I}_{2}\tilde{I}_{3}\tilde{I}_{4}}\frac{\partial X^{I_{1}}}{\partial y^{\tilde{I}_{1}}}\frac{\partial X^{I_{2}}}{\partial y^{\tilde{I}_{2}}}\frac{\partial X^{I_{3}}}{\partial y^{\tilde{I}_{3}}}\frac{\partial X^{I_{4}}}{\partial y^{\tilde{I}_{4}}})(\epsilon^{\tilde{I}'_{1}\tilde{I}'_{2}\tilde{I}'_{3}\tilde{I}'_{4}}\frac{\partial X^{I'_{1}}}{\partial y^{\tilde{I}'_{1}}}\frac{\partial X^{I'_{2}}}{\partial y^{\tilde{I}'_{2}}}\frac{\partial X^{I'_{3}}}{\partial y^{\tilde{I}'_{3}}}\frac{\partial X^{I'_{4}}}{\partial y^{\tilde{I}'_{4}}})\nonumber\\
&=&\int_{M^{10}+M^{N=3}}d^{14}x\sqrt{g}\epsilon_{I_{1}I_{2}I_{3}I_{4}I'_{1}I'_{2}I'_{3}I'_{4}I_{5}I_{6}}\epsilon^{\tilde{I}_{5}\tilde{I}_{6}}(\frac{\partial X^{I_{5}}}{\partial y^{\tilde{I}_{5}}}\frac{\partial X^{I_{6}}}{\partial y^{\tilde{I}_{6}}})\nonumber\\
&&\hskip 0.5in{}\times(\epsilon^{\tilde{I}_{1}\tilde{I}_{2}\tilde{I}_{3}\tilde{I}_{4}}\frac{\partial X^{I_{1}}}{\partial y^{\tilde{I}_{1}}}\frac{\partial X^{I_{2}}}{\partial y^{\tilde{I}_{2}}}\frac{\partial X^{I_{3}}}{\partial y^{\tilde{I}_{3}}}\frac{\partial X^{I_{4}}}{\partial y^{\tilde{I}_{4}}})(\epsilon^{\tilde{I}'_{1}\tilde{I}'_{2}\tilde{I}'_{3}\tilde{I}'_{4}}\frac{\partial X^{I'_{1}}}{\partial y^{\tilde{I}'_{1}}}\frac{\partial X^{I'_{2}}}{\partial y^{\tilde{I}'_{2}}}\frac{\partial X^{I'_{3}}}{\partial y^{\tilde{I}'_{3}}}\frac{\partial X^{I'_{4}}}{\partial y^{\tilde{I}'_{4}}})\nonumber\\
&=&\int_{M^{10}+M^{N=3}}d^{14}x\sqrt{g}\epsilon_{I_{1}I_{2}I_{3}I_{4}I'_{1}I'_{2}I'_{3}I'_{4}I_{5}I_{6}}(\epsilon^{\tilde{I}_{4}\tilde{I}_{5}}\frac{\partial X^{I_{4}}}{\partial y^{\tilde{I}_{4}}}\frac{\partial X^{I_{5}}}{\partial y^{\tilde{I}_{5}}})(\epsilon^{\tilde{I}'_{4}\tilde{I}_{6}}\frac{\partial X^{I'_{4}}}{\partial y^{\tilde{I}'_{4}}}\frac{\partial X^{I_{6}}}{\partial y^{\tilde{I}_{6}}})\nonumber\\
&&\hskip 0.5in{}\times(\epsilon^{\tilde{I}_{1}\tilde{I}_{2}}\frac{\partial X^{I_{1}}}{\partial y^{\tilde{I}_{1}}}\frac{\partial X^{I_{2}}}{\partial y^{\tilde{I}_{2}}})(\epsilon^{\tilde{I}'_{1}\tilde{I}'_{2}}\frac{\partial X^{I'_{1}}}{\partial y^{\tilde{I}'_{1}}}\frac{\partial X^{I'_{2}}}{\partial y^{\tilde{I}'_{2}}})(\epsilon^{\tilde{I}_{3}\tilde{I}'_{3}}\frac{\partial X^{I_{3}}}{\partial \tilde{I}^{I_{3}}}\frac{\partial X^{I'_{3}}}{\partial y^{\tilde{I}'_{3}}})\nonumber\\
&=&\int_{M^{10}+M^{N=3}}d^{14}x\sqrt{g}\Sigma_{n=1}^{5}\Big(\tr F^{n}-\tr R^{n}+\tr F^{n}R^{5-n}\Big)\nonumber\\
&&\hskip 0.5in{}+\delta\int_{M^{11}+M^{N=3}}d^{14}x\sqrt{g}\Big(\Sigma_{m=1}^{5}\Sigma_{n=0}^{\infty}(R)^{m}\hat{G}^{-n}+\Sigma_{m=1}^{5}\Sigma_{n=0}^{\infty}(\partial^{I}\hat{G}\partial_{I}\hat{G})^{m}\hat{G}^{-n-1}\nonumber\\
&&\hskip 1.25in{}+\Sigma_{m=1}^{5}\Sigma_{n=0}^{\infty}(\partial^{I}\phi\partial_{I}\phi)^{m}\hat{G}^{-n}+\Sigma_{m=1}^{5}\Sigma_{n=0}^{\infty}(\beta F_{IJ}F^{IJ})^{m}\hat{G}^{-n+1}+...\Big).\label{s17}
\end{eqnarray}
The first term in Eq.~(\ref{s17}) cancels the anomaly in Eq.~(\ref{s3}). However, the second term yields the action of modified gravity. In fact, we can rewrite the second term of Eq.~(\ref{s17}) for $n=m=1$ as follows:

\begin{eqnarray}
S_{\rm MOG}&=&\int_{M^{11}+M^{N=3}}d^{14}x\sqrt{g}\Big(\frac{R+\partial^{I}\phi\partial_{I}\phi}{2\hat{G}}+\frac{1}{2}(\frac{\partial_{I} \hat{G}\partial^{I} \hat{G}}{\hat{G}^{2}})+\beta F_{IJ}F^{IJ}+...\Big), \label{ps16}
\end{eqnarray}
where, using the definitions $\phi=\ln (\mu)$ and $F_{IJ}=B_{IJ}$, we obtain the following action:
\begin{eqnarray}
S_{\rm MOG}&=&\int_{M^{11}+M^{N=3}}d^{14}x\sqrt{g}\Big(\frac{R}{2\hat{G}}+\frac{\partial^{I}\mu\partial_{I}\mu}{2\hat{G}\mu^{2}}+\frac{1}{2}(\frac{\partial_{I} \hat{G}\partial^{I} \hat{G}}{\hat{G}^{2}})+\beta B_{IJ}B^{IJ}+...\Big). \label{pts16}
\end{eqnarray}
This is a version of the action of Moffat's modified gravity theory \cite{p1}, obtained from generalizing the Horava-Witten mechanism to fourteen dimensions. This model shows that the action that has been obtained by Moffat for the MOG theory is completely anomaly-free and can be applied to explain phenomenological features of the universe.

   \section{Scalar-Tensor-Vector-Fermion gravity}\label{o2}

In the previous section, we observed that by extending the Horava-Witten mechanism to fourteen dimensions, the MOG action can be obtained. However, we only studied the bosonic part of the action in supergravity, while in the Horava-Witten formalism, both fermions and bosons have been considered.  This offers the opportunity to extend the MOG action by considering the effects of fermionic gravitation fields.

\begin{figure*}[thbp]
	
	\begin{center}
		\begin{tabular}{rl}
			\includegraphics[width=8cm]{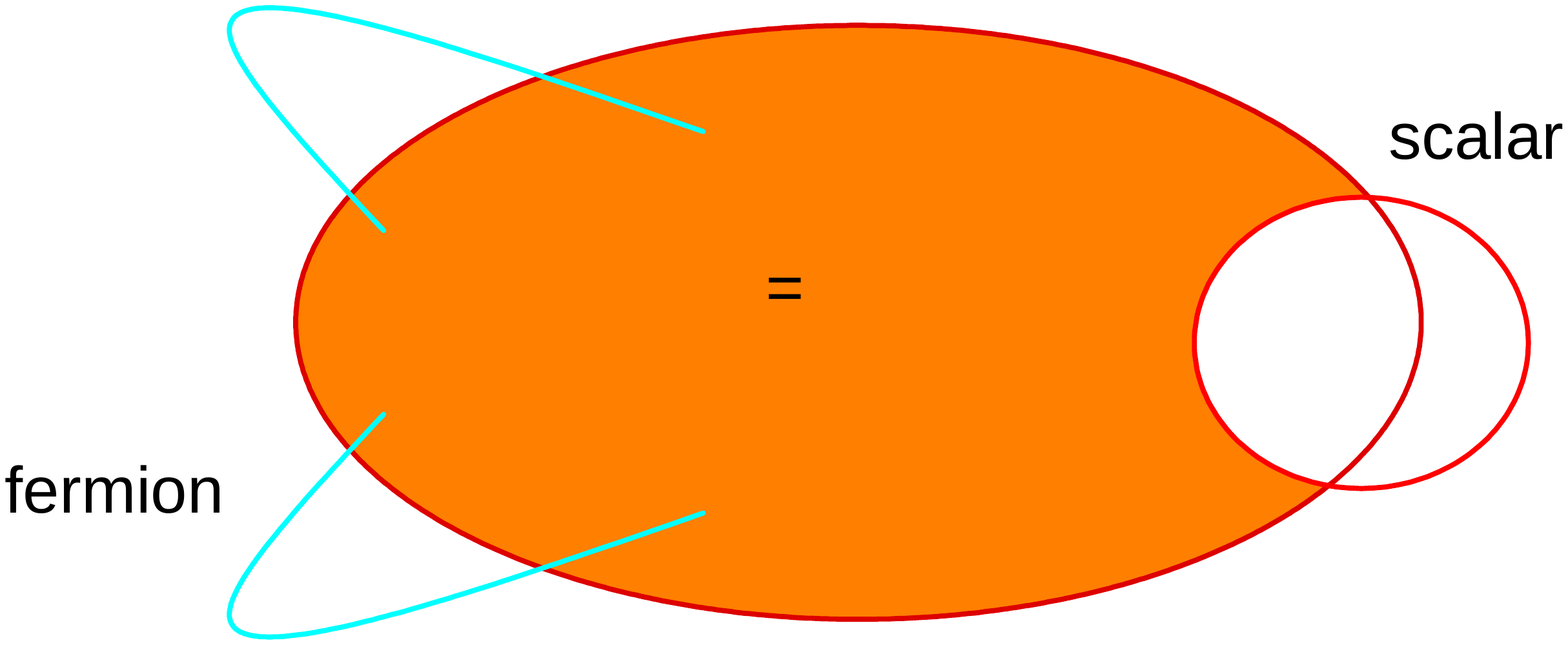}
		\end{tabular}
	\end{center}
	\caption{  Fermions can be produced by decaying scalars. }
\end{figure*}

Until now, we have shown that different shape of strings produce different types of fields. We also have argued that scalars are strings that both ends of them are located on one point like manifold. Now, if this string decays to two parts, two fermions are emerged (See Figure 13). To include fermions, we redefine strings in equation (\ref{s4}) by regarding ($\phi\rightarrow \varepsilon^{IJK}\psi^{\dag J}\psi^{K} $) and re-obtain the Poisson brackets as

\begin{eqnarray}
X^{I}&=& I^{I} + y^{I} + \epsilon^{I}\phi + A^{I} + y_{J}( F^{IJ}- R^{IJ}+\partial^{I}\phi\partial^{J}\phi-\epsilon^{IJ}\Sigma_{n}\hat{G}^{-\frac{n}{2}-1}+...) +i\varepsilon^{IJK}\psi^{\dag J}\psi^{K}\Rightarrow\nonumber\\
\{ X^{I},X^{J} \}&=&\Sigma_{I,J}\varepsilon^{I'J'}\frac{\partial X^{I}}{\partial y^{J'}}\frac{\partial X^{I}}{\partial y^{J'}}\nonumber\\
&=&\Sigma_{I,J}\varepsilon^{IJ'}\Big(\partial_{J'}A^{I}-\partial_{J'}(\varepsilon^{JK}\Gamma^{A}_{A K})\Big)\nonumber\\
&=&F^{IJ}- R^{IJ}+\partial^{I}\phi\partial^{J}\phi+i\varepsilon^{IJ'K}\psi^{\dag J'}\partial_{K}\psi^{J}-\varepsilon^{IJ}\Sigma_{n}\hat{G}^{-\frac{n}{2}-1}+...~.\label{po1}
\end{eqnarray}
where $\psi^I$ denotes a fermionic field. With this definition, the gauge variation of the $C$ term in Eq.~(\ref{s16}) can be changed to include this fermionic field:

\begin{eqnarray}
X^{I}&=& I^{I} + y^{I}+ \epsilon^{I}\phi + A^{I} + y_{J}( F^{IJ} \nonumber\\&-& R^{IJ}+\partial^{I}\phi\partial^{J}\phi-\epsilon^{IJ}\Sigma_{n}\hat{G}^{-\frac{n}{2}-1}+...) +i\varepsilon^{IJK}\psi^{\dag J}\psi^{K}\nonumber\\
&\Downarrow&\nonumber\\
\frac{\partial \delta_{A} X^{I}}{\partial y^{I}}&=&\delta ( y^{I}), \nonumber\\
&\Downarrow&\nonumber\\
\int_{M^{N=3}+M^{11}}\int d^{11}x\int dy^{I_{5}}dy^{I_{6}}dy^{I_{7}}\delta_{A} C^{I_{5}I_{6}I_{7}}&=&\int_{M^{N=2}+M^{11}}\int d^{11}x\int dy^{I_{5}}dy^{I_{6}} \Sigma_{I'_{5}I'_{6}I'_{7}}\epsilon^{I'_{5}I'_{6}I'_{7}}\delta_{A}\left(\frac{\partial X^{I_{5}}}{\partial y^{I'_{5}}}\frac{\partial X^{I_{5}}}{\partial y^{I'_{6}}}\frac{\partial X^{I_{7}}}{\partial y^{I'_{7}}}\right),\nonumber\\
&=& \int_{M^{N=2}+M^{11}}dy^{I_{5}}dy^{I_{6}}\Sigma_{I'_{5}I'_{6}}\epsilon^{I'_{5}I'_{6}}\left(\frac{\partial X^{I_{5}}}{\partial y^{I'_{5}}}\frac{\partial X^{I_{6}}}{\partial y^{I'_{6}}}\right),\nonumber\\
&=&\int_{M^{N=2}+M^{11}}dy^{I}dy^{J}(F^{IJ}- R^{IJ} \nonumber\\&+&i\varepsilon^{IJ'K}\psi^{\dag J'}\partial_{K}\psi^{J}+\partial^{I}\phi\partial^{J}\phi-\epsilon^{IJ}\Sigma_{n=1}^{\infty}\hat{G}^{-\frac{n}{2}-1})+...,\nonumber\\
&=&\int_{M^{N=2}+M^{11}}dy^{I}dy^{J}(F^{IJ}- R^{IJ}\nonumber\\&+& i\epsilon^{IJ'K}\psi^{\dag J'}\partial_{K}\psi^{J}+\partial^{I}\phi\partial^{J}\phi-\epsilon^{IJ}\hat{G}^{-1} +...).\label{po2}
\end{eqnarray}

  Also, the gauge variation of CGG terms in Eq.~(\ref{s17}) has the following shape:

\begin{eqnarray}
H^{\psi}&=&i\varepsilon^{IJK}\psi^{\dag I}\partial_{K}\psi^{J}\nonumber\\
&\Downarrow&\nonumber\\
\delta S^{N=11+3}_{\rm CGG}
&=&\delta\int_{M^{11}+M^{N=3}}d^{14}x\sqrt{g}\varepsilon_{I_{1}I_{2}I_{3}I_{4}I'_{1}I'_{2}I'_{3}I'_{4}I_{5}I_{6}I_{7}}\varepsilon^{\tilde{I}_{5}\tilde{I}_{6}\tilde{I}_{7}}\left(\frac{\partial X^{I_{5}}}{\partial y^{\tilde{I}_{5}}}\frac{\partial X^{I_{6}}}{\partial y^{\tilde{I}_{6}}}\frac{\partial X^{I_{7}}}{\partial y^{\tilde{I}_{7}}}\right) G^{I_{1}I_{2}I_{3}I_{4}}G^{I'_{1}I'_{2}I'_{3}I'_{4}}\nonumber\\
&=&\delta\int_{M^{11}+M^{N=3}}d^{14}x\sqrt{g}\varepsilon_{I_{1}I_{2}I_{3}I_{4}I'_{1}I'_{2}I'_{3}I'_{4}I_{5}I_{6}I_{7}}\varepsilon^{\tilde{I}_{5}\tilde{I}_{6}\tilde{I}_{7}}\left(\frac{\partial X^{I_{5}}}{\partial y^{\tilde{I}_{5}}}\frac{\partial X^{I_{6}}}{\partial y^{\tilde{I}_{6}}}\frac{\partial X^{I_{7}}}{\partial y^{\tilde{I}_{7}}}\right) \times \nonumber\\&&\left(\varepsilon^{\tilde{I}_{1}\tilde{I}_{2}\tilde{I}_{3}\tilde{I}_{4}}\frac{\partial X^{I_{1}}}{\partial y^{\tilde{I}_{1}}}\frac{\partial X^{I_{2}}}{\partial y^{\tilde{I}_{2}}}\frac{\partial X^{I_{3}}}{\partial y^{\tilde{I}_{3}}}\frac{\partial X^{I_{4}}}{\partial y^{\tilde{I}_{4}}}\right)\left(\varepsilon^{\tilde{I}'_{1}\tilde{I}'_{2}\tilde{I}'_{3}\tilde{I}'_{4}}\frac{\partial X^{I'_{1}}}{\partial y^{\tilde{I}'_{1}}}\frac{\partial X^{I'_{2}}}{\partial y^{\tilde{I}'_{2}}}\frac{\partial X^{I'_{3}}}{\partial y^{\tilde{I}'_{3}}}\frac{\partial X^{I'_{4}}}{\partial y^{\tilde{I}'_{4}}}\right)\nonumber\\
&=&\int_{M^{10}+M^{N=3}}d^{14}x\sqrt{g}\varepsilon_{I_{1}I_{2}I_{3}I_{4}I'_{1}I'_{2}I'_{3}I'_{4}I_{5}I_{6}}\varepsilon^{\tilde{I}_{5}\tilde{I}_{6}}\left(\frac{\partial X^{I_{5}}}{\partial y^{\tilde{I}_{5}}}\frac{\partial X^{I_{6}}}{\partial y^{\tilde{I}_{6}}}\right)\nonumber\\
&&\hskip 0.5in{}\times\left(\varepsilon^{\tilde{I}_{1}\tilde{I}_{2}\tilde{I}_{3}\tilde{I}_{4}}\frac{\partial X^{I_{1}}}{\partial y^{\tilde{I}_{1}}}\frac{\partial X^{I_{2}}}{\partial y^{\tilde{I}_{2}}}\frac{\partial X^{I_{3}}}{\partial y^{\tilde{I}_{3}}}\frac{\partial X^{I_{4}}}{\partial y^{\tilde{I}_{4}}}\right)\left(\varepsilon^{\tilde{I}'_{1}\tilde{I}'_{2}\tilde{I}'_{3}\tilde{I}'_{4}}\frac{\partial X^{I'_{1}}}{\partial y^{\tilde{I}'_{1}}}\frac{\partial X^{I'_{2}}}{\partial y^{\tilde{I}'_{2}}}\frac{\partial X^{I'_{3}}}{\partial y^{\tilde{I}'_{3}}}\frac{\partial X^{I'_{4}}}{\partial y^{\tilde{I}'_{4}}}\right)\nonumber\\
&=&\int_{M^{10}+M^{N=3}}d^{14}x\sqrt{g}\varepsilon_{I_{1}I_{2}I_{3}I_{4}I'_{1}I'_{2}I'_{3}I'_{4}I_{5}I_{6}}\left(\varepsilon^{\tilde{I}_{4}\tilde{I}_{5}}\frac{\partial X^{I_{4}}}{\partial y^{\tilde{I}_{4}}}\frac{\partial X^{I_{5}}}{\partial y^{\tilde{I}_{5}}}\right)\left(\varepsilon^{\tilde{I}'_{4}\tilde{I}_{6}}\frac{\partial X^{I'_{4}}}{\partial \tilde{I}^{I_{4}}}\frac{\partial X^{I_{6}}}{\partial y^{\tilde{I}_{6}}}\right)\nonumber\\
&&\hskip 0.5in{}\times\left(\varepsilon^{\tilde{I}_{1}\tilde{I}_{2}}\frac{\partial X^{I_{1}}}{\partial y^{\tilde{I}_{1}}}\frac{\partial X^{I_{2}}}{\partial y^{\tilde{I}_{2}}}\right)\left(\varepsilon^{\tilde{I}'_{1}\tilde{I}'_{2}}\frac{\partial X^{I'_{1}}}{\partial y^{\tilde{I}'_{1}}}\frac{\partial X^{I'_{2}}}{\partial y^{\tilde{I}'_{2}}}\right)\left(\varepsilon^{\tilde{I}_{3}\tilde{I}'_{3}}\frac{\partial X^{I_{3}}}{\partial y^{\tilde{I}_{3}}}\frac{\partial X^{I'_{3}}}{\partial y^{\tilde{I}'_{3}}}\right)\nonumber\\
&=&\int_{M^{10}+M^{N=3}}d^{14}x\sqrt{g}\Sigma_{n=1}^{5}\Big(\tr F^{n}-\tr R^{n}+\tr F^{n}R^{5-n}\Big)\nonumber\\
&+&\int_{M^{10}+M^{N=3}}d^{14}x\sqrt{g}\Sigma_{n=1}^{5}\Big(\tr(H^{\psi^{1}}...H^{\psi^{n}})+\tr(H^{\psi^{1}}...H^{\psi^{n}}(R^{5-n}+F^{5-n}))\Big)\nonumber\\
&+&\delta\int_{M^{11}+M^{N=3}}d^{14}x\sqrt{g}\Big(\Sigma_{m=1}^{5}\Sigma_{n=0}^{\infty}(R)^{m}\hat{G}^{-n}+\Sigma_{m=1}^{5}\Sigma_{n=0}^{\infty}(\partial^{I}G\partial_{I}\hat{G})^{m}\hat{G}^{-n-1}\nonumber\\
&&\hskip 1in{}+\Sigma_{m=1}^{5}\Sigma_{n=0}^{\infty}(\partial^{I}\phi\partial_{I}\phi)^{-m}\hat{G}^{-n}+\Sigma_{m=1}^{5}\Sigma_{n=0}
                        ^{\infty}(\beta F_{IJ}F^{IJ})^{m}\hat{G}^{-n+1}\nonumber\\
&&\hskip 1in{}+\Sigma_{m=1}^{5}\Sigma_{n=0}^{\infty}( H^{\psi^{1}}...H^{\psi^{m}})\hat{G}^{-n}+...\Big).\label{po3}
\end{eqnarray}

It is clear that the first two terms of Eq.~(\ref{po3}) cancel the anomaly via the Horava-Witten mechanism \cite{b1,b2}. However, the remaining terms yield the action of a modified theory of gravity that now includes a fermionic gravitational field. Putting $n=m=1$, the modified gravity action can now be written as

                           \begin{eqnarray}
                           && S_{\rm MOG}=\int_{M^{11}+M^{N=3}}d^{14}x\sqrt{g}\Big(\frac{R+i\varepsilon^{IJK}\psi^{\dag I}\partial_{K}\psi^{J}+\partial^{I}\phi\partial_{I}\phi}{2\hat{G}}+\frac{1}{2}(\frac{\partial_{I} \hat{G}\partial^{I} \hat{G}}{\hat{G}^{2}})+\beta F_{IJ}F^{IJ}+...\Big),\label{ps16f}
                           \end{eqnarray}
or, after replacing $\phi=\ln (\mu)$ and $F_{IJ}=B_{IJ}$, we obtain to following action:

                              \begin{eqnarray}
                              && S_{\rm MOG}=\int_{M^{11}+M^{N=3}}d^{14}x\sqrt{g}\Big(\frac{R}{2\hat{G}}+\frac{\partial^{I}\mu\partial_{I}\mu}{2\hat{G}\mu^{2}}+i\frac{\varepsilon^{IJK}\psi^{\dag I}\partial_{K}\psi^{J}}{2\hat{G}}+\frac{1}{2}(\frac{\partial_{I} \hat{G}\partial^{I} \hat{G}}{\hat{G}^{2}})+\beta B_{IJ}B^{IJ}+...\Big).\label{pts16f}
                              \end{eqnarray}

Also, to obtain the standard form of equation for fermions, we can use of following replacement:

    \begin{eqnarray}
	&& \psi^{J} \longrightarrow \chi \gamma^{J} \nonumber\\&& \varepsilon^{IJK}\gamma^{ I}\gamma^{J}=\gamma^{0}\gamma^{K}\nonumber\\&& \chi^{\dag}\gamma^{0} \longrightarrow \bar{\chi}\nonumber\\&&
	\varepsilon^{IJK}\psi^{\dag I}\partial_{K}\psi^{J}\longrightarrow \bar{\chi}\gamma^{K}\partial_{K}\chi\label{xs1}
\end{eqnarray}

By using above replacement, we can re-write equation (\ref{pts16f}) as:

  \begin{eqnarray}
&& S_{\rm MOG}=\int_{M^{11}+M^{N=3}}d^{14}x\sqrt{g}\Big(\frac{R}{2\hat{G}}+\frac{\partial^{I}\mu\partial_{I}\mu}{2\hat{G}\mu^{2}}+i\frac{\bar{\chi}\gamma^{K}\partial_{K}\chi}{2\hat{G}}+\frac{1}{2}(\frac{\partial_{I} \hat{G}\partial^{I} \hat{G}}{\hat{G}^{2}})+\beta B_{IJ}B^{IJ}+...\Big).\label{xs2}
\end{eqnarray}

The action of Eq.~(\ref{xs2}) is an extended version of Moffat's scalar-tensor-vector gravity theory, which now includes a fermionic gravitational field. This theory has been obtained by generalizing the Horava-Witten mechanism to fourteen dimensions and as such, it is completely anomaly-free. On the other hand, it has been shown previously that MOG fits experimental data \cite{p2,p3}. Thus, it is conceivable that our 4-dimensional universe is, in fact, part of a 14-dimensional manifold.

\section{Summary and conclusion }\label{o4}
In this paper, we obtained the exact form of the action in a Modified gravity (MOG). Our result is in agreement with a previous model that has been proposed by Moffat in \cite{p1}. Generalizing the Horava mechanism to fourteen dimensions, we demonstrated that this theory is anomaly-free.  Furthermore, we extended Moffat's action by including fermionic gravitational fields.

In next work, we can examine  Moffat's theory with  the  gravitational wave detections. Recently, it has been shown that, if advanced projects on the detection
of Gravitational Waves (GWs) will improve their sensitivity, allowing to
perform a GWs astronomy, accurate angular and frequency dependent
response functions of interferometers for GWs arising from various Theories of Gravity, i.e. General Relativity and Extended Theories of Gravity,
will be the definitive test for General Relativity \cite{co1}. Infact, Moffat's modified gravity is only an example and there are other extended theories of gravity \cite{co2,co3} which could be constructed with the mechanism of this paper and  tested by gravitational waves and the methode of ref \cite{co1}.

\section*{Acknowledgments}
\noindent The work of Alireza Sepehri has been supported
financially by Research Institute for Astronomy and Astrophysics
of Maragha (RIAAM), Iran, under research project No.1/5237-78.


\end{document}